\documentclass[%
 superscriptaddress,
 preprint,
 showpacs,preprintnumbers,
 amsmath,amssymb,
 aps,
]{revtex4}
\usepackage{graphicx}% Include figure files
\usepackage{dcolumn}% Align table columns on decimal point
\usepackage{bm}% bold math
\begin{document}

\preprint{J. Chem. Phys. 136, 000000 (2012)}
\title{Effect of glycerol and dimethyl sulfoxide on the phase behavior of lysozyme: Theory and experiments}
\author{Christoph G\"ogelein}
  \email{christoph.goegelein@ds.mpg.de}
  \affiliation{ 
    Max-Planck-Institut f\"ur Dynamik und Selbstorganisation, 
    Am Fa{\ss}berg 17, 37077 G\"ottingen, Germany
  }  
  \altaffiliation{Christoph G\"ogelein and Dana Wagner contributed equally to this work.}
\author{Dana Wagner}
  \altaffiliation{Christoph G\"ogelein and Dana Wagner contributed equally to this work.}
\author{Fr\'ed\'eric Cardinaux}
  \affiliation{ 
    Condensed Matter Physics Laboratory, Heinrich-Heine-University, 
    Universit\"atsstra{\ss}e 1, 40225 D\"usseldorf, Germany
  }
\author{Gerhard N\"agele}
  \affiliation{ 
    Institute of Complex Systems (ICS-3), Forschungszentrum J\"ulich,
    52425 J\"ulich, Germany
  }
\author{Stefan U. Egelhaaf}
  \affiliation{ 
    Condensed Matter Physics Laboratory, Heinrich-Heine-University, 
    Universit\"atsstra{\ss}e 1, 40225 D\"usseldorf, Germany
  }
\date{\today}

\begin{abstract}
Salt, glycerol and dimethyl sulfoxide (DMSO) are used to modify the properties of protein solutions. 
We experimentally determined the effect of these additives on the phase behavior of lysozyme solutions.
Upon the addition of glycerol and DMSO, the fluid-solid transition and the gas-liquid coexistence curve (binodal) shift to lower temperatures and the gap between them increases.
The experimentally observed trends are consistent with our theoretical predictions based on the thermodynamic perturbation theory (TPT) and the Derjaguin-Landau-Verwey-Overbeek (DLVO) model for the lysozyme-lysozyme pair interactions.
The values of the parameters describing the interactions, namely the refractive indices, dielectric constants, Hamaker constant and cut-off length, are extracted from literature or are experimentally determined by independent experiments, including static light scattering, to determine the second virial coefficient.
We observe that both, glycerol and DMSO, render the potential more repulsive, while sodium chloride reduces the repulsion. 
\end{abstract}


\maketitle

\section{Introduction}
\label{sec:intro}

Proteins are complex macromolecules which play a crucial role, {e.g.} as enzymes or structural units, in various processes in living organisms, but also in biotechnology.
Many of these natural or industrial processes require further ingredients and additives such as salts or solvents.
In biotechnology, these substances are often added to modify the protein interactions and phase behavior, for example to favor crystallization \cite{mcpherson:1999}. 
There is a great variety of additives; they include ions, such as sodium chloride (NaCl), to regulate the electrostatic repulsion or potential, and liquids, such as glycerol or DMSO, to, {e.g.}, stabilize proteins against denaturation \cite{sousa:1995}, protect proteins against freezing \cite{arakawa:2007,kamiyama:2009} or inhibit protein aggregation \cite{mcpherson:1990}.
A thorough understanding of the resulting changes in the interactions and phase behavior is beneficial for the design and control of these processes.
For example, experience has shown that high quality crystals are often formed in slightly supersaturated solutions \cite{mcpherson:1999}, while protein crystals with many defects or amorphous protein precipitates (aggregates) are frequently obtained if the sample is quenched deep into the crystal phase, in particular close to the metastable gas-liquid coexistence curve where rapid nucleation occurs \cite{wolde:1997,sedgwick:2005}.
Thus, understanding the phase behavior can guide the design of processes involving proteins.

Despite the huge complexity of proteins, even of globular proteins such as lysozyme considered here, we will use a very simple spherically symmetric model.
Although this will not do full justice to the details of the protein structure and the individual properties of a specific protein, it will allow us to apply concepts developed in soft matter physics to protein solutions. 
It will also shed light on general properties of proteins which can successfully be described by a coarse-grained model.
On the other hand, the model will fail to describe certain aspects and thus indicate what defines the specific properties of a certain protein. 
To describe the protein-protein interactions, we use the Derjaguin-Landau-Verwey-Overbeek (DLVO) potential, a well-established model potential in colloid physics \cite{israelachvili:1991,pusey:1991}. 
Based on this interaction model and thermodynamic perturbation theory (TPT) \cite{barker:1976}, we will calculate the phase behavior of lysozyme in aqueous solution with additives, namely added salt (NaCl), glycerol and dimethyl sulfoxide (DMSO), and compare our theoretical predictions to our experimental observations. 
For the additive concentrations studied here, lysozyme does not unfold or undergo conformational changes \cite{knubovets:1999,vagenende:2009,voets:2010}.
This allows us to test the validity and limits of the simple model, to investigate the effects of these additives on the interactions and the phase behavior, and to understand at least qualitatively the trends observed in our experiments.

In several previous studies, the properties of lysozyme solutions, namely the protein-protein interactions and the phase behavior, were described using the DLVO potential. 
Poon {\it et al.}~\cite{poon:2000} studied the fluid-solid transition, i.e.~the crystallization or solubility boundary, as a function of salt and protein concentration as well as the pH of the solution and thus the protein charge. 
They found a universal crystallization boundary if the salt concentration is normalized by the square of the protein charge and supported this finding by the behavior of the second virial coefficient calculated using the DLVO potential. 
A more detailed analysis was later provided by Warren \cite{warren:2002}. 
Similarly, computer simulations by Pellicane {\it et al.}~\cite{pellicane:2003} based on the DLVO model showed that the shape of the gas-liquid coexistence curve can successfully be predicted, with the critical temperature strongly depending on the Hamaker constant. 
By contrast, Broide {\it et al.}~\cite{broide:1996} report that the DLVO model cannot fully explain their experimentally-observed phase behavior; the predicted height of the Coulomb barrier is too low to inhibit protein aggregation. 
This suggests that hydration forces are important to prevent lysozyme from aggregating.

Despite the widespread use of additives such as glycerol and DMSO, for example in crystallization essays, only a few studies were concerned with their effect on the phase behavior of proteins.
Farnum and Zukoski \cite{farnum:1999} studied the effect of glycerol on the protein interactions in aqueous solutions of bovine pancreatic trypsin inhibitor (BPTI). 
Upon increasing the glycerol concentration, the repulsive interaction increases. 
A description of these data based on the DLVO model requires a Hamaker constant which is much smaller than expected from the optical properties of water-glycerol mixtures and of proteins. 
They thus attributed their observations to a small increase of the effective protein size which is caused by an enhanced hydration shell (and modeled by an increased cut-off length $\delta$, see Sec.~\ref{sec:values}). 
This is supported by observations of Priev {\it et al.}~\cite{priev:1996}. 
Their experiments indicate an increase of the hydration shell upon addition of glycerol and, at the same time, a decrease of the core volume of BPTI. 
Similarly, using small-angle neutron scattering Sinibaldi {\it et al.}~\cite{sinibaldi:2007} observed, upon increasing the glycerol content, a significant increase of the hydration shell of lysozyme from $0.3\,\mathrm{nm}$ to $0.6\,\mathrm{nm}$, and a small decrease of the core volume of lysozyme of about $6\%$, such that the total effective protein volume increases by at least about $13\%$ \cite{merzel:2002, sinibaldi:2007}.
In the presence of glycerol, Esposito {\it et al.}~\cite{esposito:2009} furthermore observed an increase of the hydrodynamic radius with increasing temperature. 
Using computer simulations, Vagenende {\it et al.}~\cite{vagenende:2009} found that glycerol preferentially interacts with the hydrophobic patches on the lysozyme surface.
Glycerol molecules shield these patches with their hydrocarbon backbone, allowing the lysozyme hydration shell to grow.
Moreover, Sedgwick {\it et al.}~\cite{sedgwick:2007} studied the effect of up to $40\,\mathrm{vol.}\%$ glycerol on the phase behavior of lysozyme. 
They found that, based on the DLVO model, the phase behavior can be described almost quantitatively if the glycerol-induced changes in the dielectric constant and index of refraction are taken into account.
The effect of glycerol and DMSO on the fluid-solid boundary of lysozyme solutions has also been studied by Lu {\it et al.}~\cite{lu:2003}. 
Both additives increase the lysozyme solubility. 
Furthermore, consistent with the previously-mentioned study \cite{sedgwick:2007}, Lu {\it et al.}~\cite{lu:2003} found that glycerol and DMSO decrease the protein critical supersaturation, thereby promoting nucleation.
In subsequent work, Gosavi {\it et al.} \cite{gosavi:2009} studied the formation of lysozyme crystals showing that the crystal growth rate is enhanced upon addition of glycerol. 
Similar observations were made by Kulkarni and Zukoski \cite{kulkarni:2002}.
Furthermore, Arakawa {\it et al.}~\cite{arakawa:2007} and Kamiyama {\it et al.}~\cite{kamiyama:2009} observed a preferential hydration of lysozyme in water-DMSO mixtures, indicating that DMSO does not bind to the lysozyme surface  \cite{kamiyama:2009}. 
This suggests that DMSO mainly affects the dielectric properties of the bulk solution. 
Nevertheless, DMSO leads to denaturation at volume fractions beyond $70\,\mathrm{vol.}\%$ \cite{voets:2010}.
It is hence still not clear whether the addition of glycerol and DMSO only changes the solvent dielectric constant and refractive index or whether it also has a significant effect on the protein, in particular on its hydration.

Combining theory and experiment, we investigate the phase behavior of aqueous lysozyme solutions with additives. 
The samples contain  $0.7\,\mathrm{M}$ or $0.9\,\mathrm{M}$ NaCl, up to $20\,\mathrm{vol.}\%$ glycerol and up to $15\,\mathrm{vol.}\%$ DMSO.
By optical inspection, we experimentally determine both the fluid-solid transition and the metastable gas-liquid coexistence curves (binodals). 
Of particular interest is their relative location in dependence on the amount of added glycerol and DMSO, since protein crystallization is expected to be enhanced in the vicinity of the gas-liquid critical point \cite{wolde:1997}. 
As we will show, the experimentally-determined phase diagram of lysozyme is qualitatively consistent with our theoretical predictions.
They are based on the DLVO model and thermodynamic perturbation theory (TPT) \cite{barker:1976}, and are preformed without any free parameters. 
The values of the parameters describing the DLVO model are either taken from the existing literature or determined by independent measurements, namely refractometry to measure the refractive indices and static light scattering (SLS) to determine the second virial coefficient. 
These independent measurements also provide detailed information on the effect of the additives on the protein interactions.

The organization of this paper is as follows.
In Section \ref{experiments}, we describe the experimental procedures.
Section \ref{sec:model} introduces the DLVO model for the protein-protein interactions and presents the values of its parameters. 
Based on this model and without additional free parameters, in Section~\ref{sec:phase} the phase behavior is calculated using thermodynamic perturbation theory.
Then the predicted phase behavior is compared to our experimental observations.
Finally, our findings are summarized in Section \ref{conclusion}.

%%%%% Materials and Experimental Methods %%%%%

\section{\label{experiments}Materials and Experimental Methods}

\subsection{Materials and Sample Preparation}
\label{sec:materials}

We used three-times crystallized, dialyzed and lyophilized hen egg-white lysozyme powder (Sigma Aldrich L6876).
An aqueous $50\,\mathrm{mM}$ sodium acetate buffer was prepared and adjusted with hydrochloric acid to $p\mathrm{H}=4.5$. 
The protein powder was dissolved in this buffer to result in a concentration of about $40\,\mathrm{mg/ml}$.
This suspension was several times passed through a filter with pore size $0.1\,\mu\mathrm{m}$ (Acrodisc syringe filter, low protein binding, Pall 4611) to remove impurities and undissolved protein.
This solution subsequently was concentrated by a factor of six using an Amicon stirred ultra-filtration cell (Amicon, Millipore, 5121) with an Omega 10k membrane disc, Pall OM010025. 
The concentrated solution was used as protein stock solution.
Its protein concentration was determined in a quartz cell with a path length of $1\,\mathrm{cm}$ by UV absorption spectroscopy at a wavelength of $280\,\mathrm{nm}$ using a specific absorption coefficient of $E=2.64\,\mathrm{ml/(mg\, cm)}$.

Samples were prepared by mixing appropriate amounts of protein stock solution, buffer, salt solution ($3\,\mathrm{M}$ NaCl with buffer), glycerol solution ($70\,\mathrm{vol.}\%$ glycerol with buffer), and DMSO. 
Mixing was performed at a temperature above the metastable gas-liquid coexistence to prevent phase separation.
The glycerol samples contained $0.9\,\mathrm{M}$ and $0.7\,\mathrm{M}$ NaCl, the DMSO samples $0.7\,\mathrm{M}$ NaCl, and all samples contained buffer and various concentrations of lysozyme.

In order to obtain protein concentrations above that of the stock solution, we exploited the gas-liquid phase separation.
A solution was quenched into the metastable gas-liquid coexistence region, centrifuged for 10 min at a temperature below the cloud point temperature, and the protein-rich phase was collected.
The resulting protein concentration of the protein-rich phase was calculated from the volume ratio of the two coexisting phases, the initial protein concentration and the protein concentration of the protein-poor phase, which was determined via its refractive index.

\subsection{Determination of the Phase Boundaries}

The gas-liquid coexistence curves (binodals) were determined by cloud point measurements.
The sample was filled into a NMR tube with $5\,\mathrm{mm}$ diameter, sealed, and placed into a water bath at a temperature above the demixing point.
The temperature subsequently was lowered stepwise and the cloud point temperature identified by the sample becoming turbid.
This measurement was repeated for samples with different lysozyme volume fractions to determine the location of the binodals. 

To determine the crystallization boundary, the sample was filled into an x-ray capillary tube and sealed with UV curing glue (Norland). 
The capillary was then mounted onto a home-built temperature stage that allows for observation by optical microscopy (Nikon Eclipse 80i).
For samples with a low protein concentration, it was necessary to lower the temperature to $+4^{\circ}\mathrm{C}$ to induce crystallization.
Once crystals were formed, the temperature was raised stepwise.
The temperature at which the crystals begin to melt was identified with the transition temperature.
Again, the experiments were repeated with different lysozyme volume fractions to obtain the fluid-solid coexistence curves.

\subsection{Determination of the Refractive Indices}
\label{sec:refractometer}

We used a temperature-controlled Abbe refractometer (Model 60L/R, Bellingham \& Stanley) operated at a wavelength of $589.6\,\mathrm{nm}$ to determine the refractive indices of the samples, that is of the protein solutions, $n$, and of the solvent, $n_{\mathrm{s}}$.
The refractive index increments, $\mathrm{d}n/\mathrm{d}c_{\mathrm{p}}$, were obtained from linear fits to the dependence of $n$ on the mass protein concentration, $c_{\mathrm{p}}$.

\subsection{Determination of the Second Virial Coefficient}
\label{sec:B2}

Static light scattering (SLS) was performed with a 3D light scattering instrument (LS-Instruments).
Due to the low scattering intensity of the samples, the instrument was operated with a single beam with wavelength $\lambda=633\,\mathrm{nm}$.
The samples were filled into cylindrical glass cuvettes with a diameter of $10\,\mathrm{mm}$, centrifuged for at least 10~min at typically $7\,500\,\mathrm{g}$ prior to the measurements, and placed into the temperature-controlled vat of the instrument which was filled with decalin.   
The samples used for the SLS measurements were dilute with lysozyme concentrations between $2.8$~mg/ml and  $14.4$~mg/ml.

We did not observe a dependence of the scattered intensity, $I_{\mathrm{p}}$, on the scattering angle, $\theta$, consistent with the small protein diameter $\sigma=3.4\,\mathrm{nm}$ \cite{muschol:1995,kuehner:1997} compared to the inverse of the scattering vector $q=(4\pi n /\lambda) \sin{(\theta/2)}$ so that $q\,\sigma \ll 1$ \cite{kerker:1969}. 
Therefore, the time-averaged intensity scattered by the protein solution, $\langle I_{\mathrm{p}}(c_{\mathrm{p}}) \rangle$, was detected at a single scattering angle, $\theta=90^{\circ}$, corresponding to a magnitude of the scattering vector $q \approx 0.018\,\mathrm{nm}^{-1}$.  
The excess scattering due to the protein is expressed as the Rayleigh ratio,  
\begin{equation}\label{rayleigh_ratio}
  R(c_{\mathrm{p}}) = \frac{\langle I_{\mathrm{p}}(c_{\mathrm{p}})\rangle - \langle I_{\mathrm{s}}\rangle}{\langle I_{\mathrm{t}}\rangle}\,\,\frac{n^{2}}{n^{2}_{\mathrm{t}}}\,\,R_{\mathrm{t}}\,\mbox{,} 
\end{equation}
which relates the difference in the average scattering intensities of the protein solution and the solvent, $\langle I_{\mathrm{p}}(c_{\mathrm{p}})\rangle - \langle I_{\mathrm{s}}\rangle$, to the scattered intensity of the toluene reference,  $\langle I_{t}\rangle$, with known Rayleight ratio, $R_{t}$.  
The ratio of the refractive indices of the protein solution, $n$, and toluene, $n_{\mathrm{t}}$, accounts for the different sizes of the scattering volumes in the two cases.
The temperature dependence of $R_{t}$ was experimentally determined from the temperature dependence of the intensity scattered by a toluene sample.
The absolute value $R_{\mathrm{t}}=1.40\times{10}^{-5}{\mathrm{cm}}^{-1}$ at $T=35^\circ$C was taken from literature \cite{narayanan:2003} and is consistent with other reports (see \cite{itakura:2006} and references therein).

For dilute solutions, $R(c_{\mathrm{p}})$ is related to the second virial coefficient, $B_2$, by
\begin{equation}\label{rayleigh_ratio_b2_relation}
  \frac{K\,c_{\mathrm{p}}}{R(c_{\mathrm{p}})} = \frac{1}{M^{(0)}} \left (1+ \frac{2\, N_{\mathrm{A}}\, B_{2}}{M^{(1)}}\,\,c_{\mathrm{p}} \right )\,\mbox{,}  
\end{equation}
where $M^{(0)}$ and $M^{(1)}$ both represent the molar mass $M$, but by two different values (see below), $N_{\mathrm{A}}$ is Avogadro's constant and $K$ an optical constant given by
\begin{equation}\label{optical_constant}
  K = \frac{4 \pi^2 n^2_{\mathrm{s}}}{N_{\mathrm{A}} \lambda^4}\,\left(\frac{\mathrm{d}n}{\mathrm{d}c_{\mathrm{p}}}\right)^2\mbox{.}
\end{equation}
\noindent Note that the refractive indices and the refractive index increments in the above equations depend on temperature. 

SLS measurements of a series of protein concentrations hence allow to determine $B_{2}$ from the slope of the $c_{\mathrm{p}}$ dependence, and the molar mass $M$ from the intercept at $c_{\mathrm{p}}\to 0$. 
The measurements were performed at different temperatures $T$ to obtain the temperature dependence of $B_{2}(T)$, while $M$ is expected to be temperature independent. 
The values of the molar mass, experimentally determined at different temperatures from the intercept at $c_{p}\to 0$, lie in the range $14\,800$~g/mol~$\lesssim M^{(0)} \lesssim 19\,500$~g/mol, while the expected value is $14\,320\,\mathrm{g/mol}$ \cite{voet:1990}.
The error in $M^{(0)}$ is mainly due to the low scattered intensity, uncertainties in ${\mathrm{d}}n/{\mathrm{d}}c_{\mathrm{p}}$ and $R_{\mathrm{t}}$. 
Dynamic light scattering experiments on the same dilute samples did not indicate the presence of aggregates, which are also neither expected for aqueous
solutions \cite{stradner:2004,sedgwick:2005} nor for glycerol-water \cite{vagenende:2009} or DMSO-water mixtures \cite{voets:2010}. 
To avoid that experimental errors in the molar mass affect $B_{2}(T)$, we used the literature value $14\,320\,\mathrm{g/mol}$ in the term for the slope, $2N_{\mathrm{A}}B_2/M^{(1)}$, {i.e.} $M^{(1)}=14\,320\,\mathrm{g/mol}$, when calculating $B_{2}(T)$ based on {Eq.~(\ref{rayleigh_ratio_b2_relation}}) \cite{comment_1}.
This renders $B_{2}(T)$ independent of the absolute scattered intensity, {i.e.} $M^{(0)}$, which has a significant uncertainty, and thus makes $B_{2}(T)$ more reliable.

%%%%% Results and Discussion %%%%%

\section{\label{theory}Results and Discussion}

We first present the simple model used to quantify the protein-protein interactions, namely the DLVO potential model (Sec.~\ref{sec:DLVO}), as well as the values of its parameters.
These are either taken from the literature, or determined by independent measurements (Sec.~\ref{sec:values}). 
Based on this interaction potential and these parameter values, we calculate the phase diagram without any free parameter (Sec.~\ref{TPT}). 
Our experimentally determined phase diagrams (Sec.~\ref{results_and_discussion}) are then compared to the theoretical predictions (Sec.~\ref{sec:comp}).

%%%%% Protein-Protein Interaction Potential

\subsection{Protein-Protein Interaction Potential}
\label{sec:model}

\subsubsection{\label{model}DLVO Model}
\label{sec:DLVO}

The DLVO pair potential, $u(r)$, includes three parts; hard-sphere $u_{0}(r)$, repulsive electrostatic $u_{\mathrm{el}}(r)$ and attractive van der Waals interactions $u_{\mathrm{vdW}}(r)$,
\begin{equation}\label{total_pair_potential}
  u(r) = u_{0}(r) + u_{\mathrm{el}}(r) + u_{\mathrm{vdW}}(r)\,\mbox{,}
\end{equation}
where $r$ is the center-to-center distance of two spherical particles. 

\noindent The hard-sphere contribution is
\begin{equation}\label{hard_sphere_potential}
  u_{0}(r) = \left\{
                          \begin{array}{ll}
                             \infty\,\mbox{,} & r < \sigma\\
                             0\,\mbox{,} & r \geq \sigma\,\mbox{.}
                          \end{array}
                       \right. 
\end{equation}
\noindent The electrostatic repulsion is described by a screened Coulomb potential
\begin{equation}\label{beta_u_el}
  \beta u_{\mathrm{el}}(r) = \frac{Z^2 l_{\mathrm{B}}}{{(1+\kappa\sigma/2)}^2}\, \frac{\exp{\left[-\kappa(r-\sigma)\right]}}{r}\quad ( r \geq \sigma)\,\mbox{,}
\end{equation}
\noindent where $\beta=1/(k_{\mathrm{B}}T)$ with $k_{\mathrm{B}}$ the Boltzmann constant, $\sigma$ the diameter, $Z$ the charge and $l_{\mathrm{B}}=$ $e^2/$ $(4\pi\varepsilon_{0}\varepsilon_{\mathrm{s}} k_{\mathrm{B}}T)$ the solvent-specific Bjerrum length depending on the elementary charge, $e$, dielectric constant {\it in vacuo}, $\varepsilon_{0}$, and dielectric constant of the solvent, $\varepsilon_{s}$.
The square of the electrostatic (Debye) screening parameter, $\kappa$, is given by
\begin{equation}\label{debye_parameter}
  \kappa^2 = \frac{4\pi l_{\mathrm{B}}N_{\mathrm{A}}}{1-\phi}\left( Z \frac{{\rho}_{\mathrm{p}}}{M} \phi + 2 c_{\mathrm{s}} + 2 c_{\mathrm{b}} \right)\,\mbox{,}
\end{equation} 
where $\phi=c_{\mathrm{p}}/\rho_{\mathrm{p}}$ is the particle volume fraction with $\rho_{\mathrm{p}}$ the mass density, $c_{\mathrm{s}}$ the molar salt concentration and $c_{\mathrm{b}}$ the molar concentration of dissociated buffer.
The factor $1/(1-\phi)$ corrects for the volume occupied by proteins and thus takes into account the free volume accessible to the micro-ions, which cannot penetrate the protein \cite{russel:1981,denton:2000}. 
We ignore here the small free volume corrections arising from the finite size of the microions.

The van der Waals attraction is of the form \cite{israelachvili:1991}
\begin{equation}\label{eq_vdW_interaction}
u_{\mathrm{vdW}}(r)= -\frac{A}{12}\left(\frac{\sigma^2}{r^2-\sigma^2}+\frac{\sigma^2}{r^2}+2\ln\left[1-\frac{\sigma^2}{r^2}\right]\right)\quad ( r \geq \sigma + \delta)\,\mbox{,} 
\end{equation}
\noindent where $A$ is the (effective) Hamaker constant and $\delta$ a cut-off length. 
The cut-off length $\delta$ represents the smallest possible surface-to-surface separation of two proteins, introduced to avoid the unphysical $1/(r-\sigma)$ divergence at $r\to \sigma$.
Its value can be estimated by the thickness of the Stern layer, {i.e.}, by the mean size of the salt ions.

The Hamaker constant can be approximated by \cite{israelachvili:1991}
\begin{equation}\label{Hamaker_constant}
A = A_{\nu = 0} + A_{\nu > 0} = \frac{3}{4}k_{\mathrm{B}}T{\left(\frac{\varepsilon_{\mathrm{p}}-\varepsilon_{\mathrm{s}}}{\varepsilon_{\mathrm{p}}+\varepsilon_{\mathrm{s}}}\right)}^2 + \frac{3 h {\nu}_{\mathrm{e}}}{16\sqrt{2}}\,\frac{{(n_{\mathrm{p}}^{2}-n_{\mathrm{s}}^{2})}^2}{{(n_{\mathrm{p}}^2+n_{\mathrm{s}}^2)}^{3/2}}\,\mbox{,}
\end{equation} 
\noindent where $A_{\nu  = 0}$ is the zero-frequency and $A_{\nu > 0}$ the frequency-dependent part, $h$ the Planck constant, $\varepsilon_{\mathrm{s}}$ and $\varepsilon_{\mathrm{p}}$ the static dielectric constants of the solvent and protein, $n_{\mathrm{s}}$ and $n_{\mathrm{p}}$ their refractive indices, and $\nu_{e}=3\times 10^{15}\,s^{-1}$ the characteristic UV electronic adsorption frequency \cite{israelachvili:1991}.

\subsubsection{Values of the Model Parameters}
\label{sec:values}

To quantitatively determine the potential $u(r)$, we require the values of the refractive indices and static dielectric constants, of the cut-off length $\delta$, and of the parameters describing the protein.
Lysozyme is an approximately spherical protein with an effective diameter $\sigma=3.4\,\mathrm{nm}$ \cite{muschol:1995,kuehner:1997}, molar mass $M=14\,320\,\mathrm{g/mol}$, mass density $\rho_{\mathrm{p}}=1.351\,\mathrm{g/cm^3}$ \cite{voet:1990} and $Z=11.4$ net positive elementary charges at the present $p\mathrm{H}=4.5$ \cite{tanford:1972}.
The acidity constant of sodium acetate is $p\mathrm{K}_{a}=4.76$ resulting in a dissociated buffer concentration of $c_{b}=27\,\mathrm{mM}$ at $p\mathrm{H}=4.5$.

\bigskip

\paragraph{Refractive indices and static dielectric constants.}
The protein index of refraction is obtained by the linear extrapolation, $n_{p}=6\,M/(N_{\mathrm{A}}\pi\sigma^3)\;\mathrm{d}n/\mathrm{d}c_{\mathrm{p}}+n_{\mathrm{s}}$, where $6/(N_{\mathrm{A}}\pi\sigma^3)$ is the molar concentration of a `bulk' protein. 
We find no significant dependence of $\mathrm{d}n/\mathrm{d}c_{\mathrm{p}}$ on composition or temperature in the relevant parameter range, consistent with previous findings \cite{Yin:2003}. 
Averaging $\mathrm{d}n/\mathrm{d}c_{\mathrm{p}}$ over all tested temperatures and compositions yields $n_{\mathrm{p}}=1.72$.

It is usually assumed that dry protein has a static dielectric constant in the range of $2 \le \varepsilon_{\mathrm{p}} \le 4$ \cite{dwyer:2000, sedgwick:2007}. 
However, measurements with protein powders often indicate larger values, for example $1 \le \varepsilon_{\mathrm{p}} \le 10$ for wet lysozyme powder \cite{harvey:1972}, $20 \le \varepsilon_{\mathrm{p}} \le 37$ for hydrated lysozyme crystals \cite{rashkovich:2008} and, based on MD simulations, $\varepsilon_{\mathrm{p}} = 25.7$ for lysozyme in aqueous solution \cite{pitera:2001}. 
The larger values are attributed to water or moisture leading to hydration and some water penetration \cite{dwyer:2000}. 
For consistency with previous work \cite{dwyer:2000, sedgwick:2007}, we use a static dielectric constant of lysozyme given by $\varepsilon_{\mathrm{p}}=2$. 
We might underestimate in this way $\varepsilon_{\mathrm{p}}$ and thus overestimate the static contribution to the Hamaker constant $A_{\nu = 0}$ (Eq.~\ref{Hamaker_constant}) and hence the protein attraction. 
However, the effects are not significant, because, as will be shown below, the main contribution to the Hamaker constant is due to the non-zero frequency term $A_{\nu > 0}$.

%--------------------------------------------------------------------------
\begin{figure}[t!]
\includegraphics[width=0.9\textwidth]{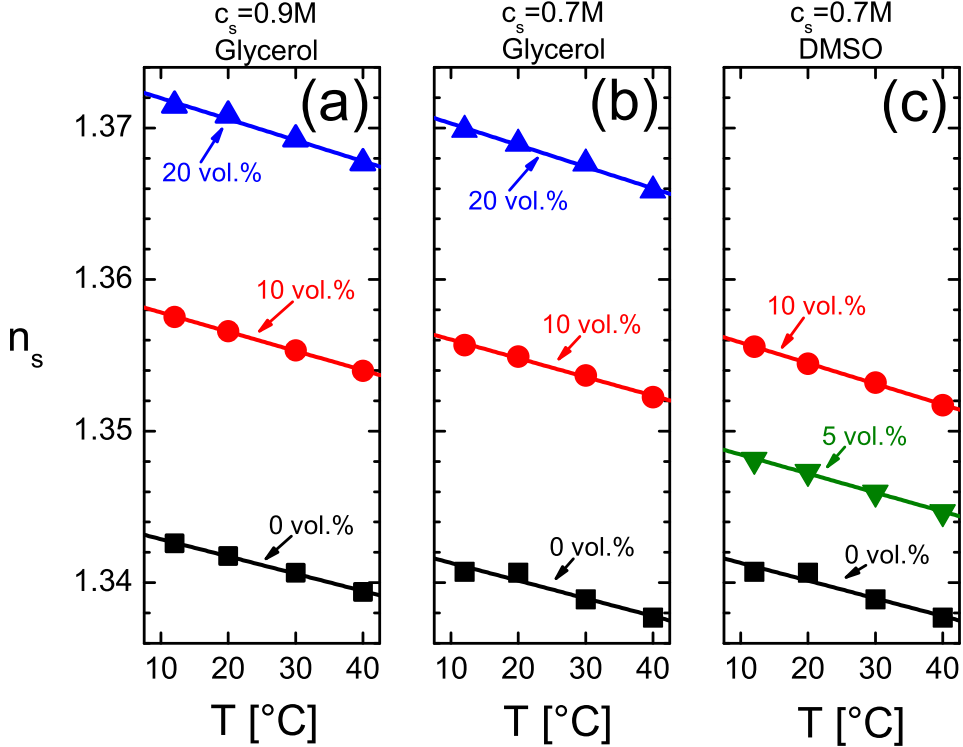}
\caption{\label{refractive_index_solvent}Temperature dependence of the index of refraction of the solvent, $n_{\mathrm{s}}(T)$, for different added NaCl concentrations, $c_{\mathrm{s}}$, as well as glycerol and DMSO contents (as indicated). The lines are linear fits to the experimental data points.}
\end{figure}
%--------------------------------------------------------------------------

The refractive index of the solvent, $n_{s}$, was measured and found to decrease with increasing temperature (Fig.~\ref{refractive_index_solvent}). 
The refractive indices of glycerol, $n_{\mathrm{g}}$, and DMSO, $n_{\mathrm{DMSO}}$, are larger than the one of water, $n_{\mathrm{w}}$.
For example, $n_{\mathrm{g}}=1.47$, $n_{\mathrm{DMSO}}=1.48$ and $n_{\mathrm{w}}=1.33$ at $T=20^{\circ}C$. 
The addition of glycerol or DMSO thus increases $n_{\mathrm{s}}$.

%--------------------------------------------------------------------------
\begin{figure}[t!]
\includegraphics[width=0.9\textwidth]{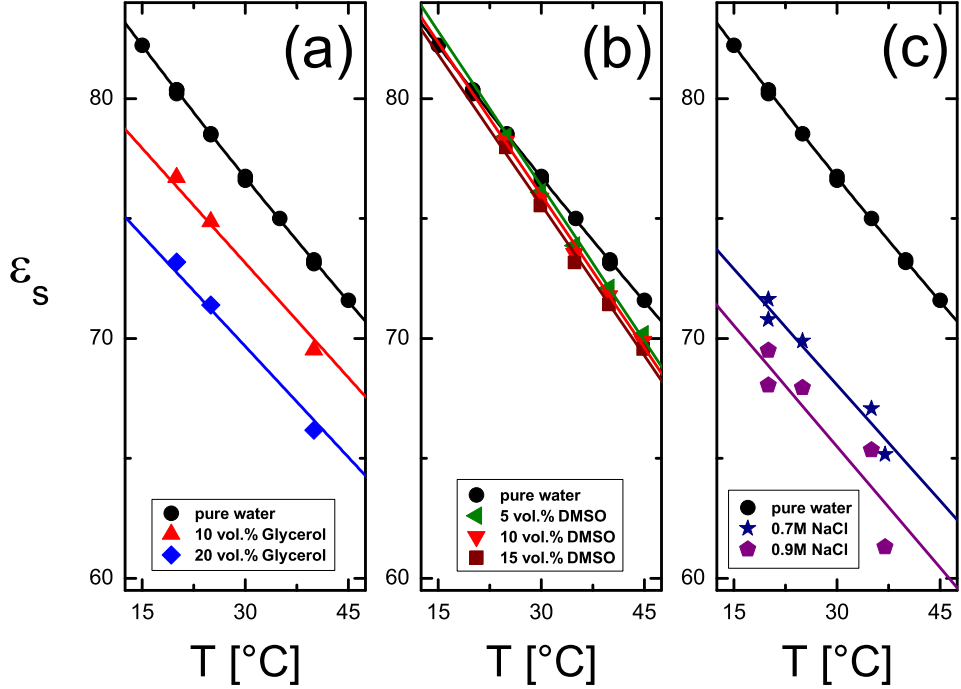}
\caption{\label{dielectric_constant}Temperature dependence of the dielectric constant, $\varepsilon_{\mathrm{s}}(T)$, of different solvent mixtures. (a) water-glycerol \cite{lide:2005,dow:2010,tpub:2010}, (b) water-DMSO \cite{lu:2009}, (c) aqueous salt solutions \cite{nortemann:1997,buchner:1999,gulich:2009}. The lines are linear fits. The fits that belong to solutions containing $10$ and $20\,\mathrm{vol.}\%$ glycerol are based on five data points in an extended temperature range $20^{\circ}\mathrm{C} \le T \le 80^{\circ}\mathrm{C}$.}
\end{figure}
%--------------------------------------------------------------------------

The dielectric constant of the solvent, $\varepsilon_{s}$, was taken from literature \cite{lide:2005,dow:2010,tpub:2010,lu:2009,nortemann:1997,buchner:1999,gulich:2009} (Fig.~\ref{dielectric_constant}). 
It decreases with increasing temperature, which is due to the increasing thermal motion of the solvent molecules and reduces the alignment of the dipolar molecules by an external electric field.
The dielectric constants of glycerol, $\varepsilon_{\mathrm{g}}$, and DMSO, $\varepsilon_{\mathrm{DMSO}}$, are much smaller than the one of water, $\varepsilon_{\mathrm{w}}$, for example $\varepsilon_{\mathrm{g}}=41$, $\varepsilon_{\mathrm{DMSO}}=48$ and $\varepsilon_{\mathrm{w}}=80$ at $T=20^{\circ}\mathrm{C}$. 
Therefore, the solvent dielectric constant, $\varepsilon_{\mathrm{s}}$, decreases with increasing glycerol and DMSO content, with the decrease being more pronounced for glycerol (Figs.~\ref{dielectric_constant}a and b). 
There is also a significant decrease of $\varepsilon_{\mathrm{s}}$ with increasing salt concentration, $c_{\mathrm{s}}$; adding $0.7\,\mathrm{M}$ NaCl lowers $\varepsilon_{s}$ more than adding $20\,\mathrm{vol.}\%$ glycerol (Fig.~\ref{dielectric_constant}c) \cite{nortemann:1997,buchner:1999,gulich:2009}. 
This decrease is related to the decrease in the electric polarizability of the solution caused by the alignment and attraction of the water molecules by the salt ions.

%--------------------------------------------------------------------------
\begin{figure}[t!]
\includegraphics{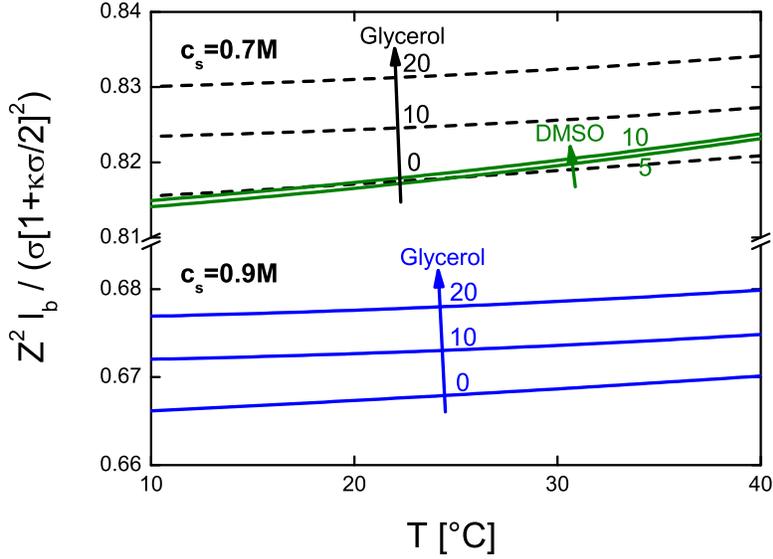}
\caption{\label{plot_contact_value_Coulomb_repulsion}Temperature dependence of the contact value of the reduced electrostatic interaction potential, $\beta u_{\mathrm{el}}(\sigma)=Z^2l_{\mathrm{B}}/(\sigma[1+\kappa \sigma /2]^2)$, for different added NaCl concentrations, $c_{\mathrm{s}}$, as well as glycerol and DMSO volume fractions (in $\mathrm{vol.}\%$, as indicated). Note the break in the ordinate axis.}  
\end{figure}
%--------------------------------------------------------------------------

Based on these static dielectric constants, we calculate the repulsive electrostatic interaction part, $u_{\mathrm{el}}(r)$, ({Eqs.~(\ref{beta_u_el}) and (\ref{debye_parameter})}) for the different compositions, i.e.~different added salt, glycerol and DMSO content. 
Upon increasing the salt (NaCl) concentration $c_{\mathrm{s}}$, the screening parameter, $\kappa$, increases and thus the contact value of the electrostatic potential, $u_{\mathrm{el}}(\sigma)$, decreases ({Fig.~\ref{plot_contact_value_Coulomb_repulsion}}). 
In addition, $u_{\mathrm{el}}(\sigma)$ depends on the solution composition since the Bjerrum length $l_{\mathrm{B}}$ depends on $\varepsilon_{s}$. 
Adding glycerol or DMSO decreases $\varepsilon_{s}$ ({Figs.~\ref{dielectric_constant}}a and b) and hence increases $u_{\mathrm{el}}(\sigma)$, with the effect of glycerol being significantly stronger.

%--------------------------------------------------------------------------
\begin{figure}[t!]
\includegraphics[width=0.9\textwidth]{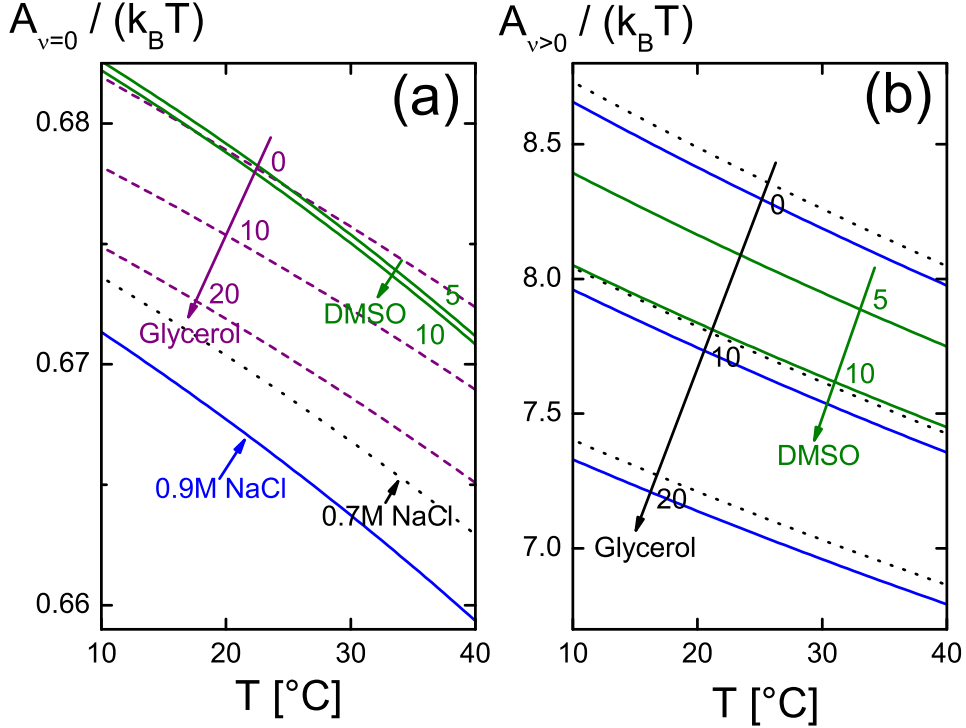}
\caption{\label{plot_Hamaker_constant}Temperature dependence of the (a) zero-frequency, $A_{\nu=0}$, and (b) non-zero frequency, $A_{\nu>0}$, parts of the Hamaker constant in {Eq.~(\ref{Hamaker_constant})} for different compositions. (a) water-glycerol mixtures without NaCl (purple dashed lines), water-DMSO mixtures without NaCl (green solid lines), aqueous salt solutions with 0.7~M NaCl (black dotted line) and 0.9~M NaCl (blue solid line), (b) water-glycerol mixtures containing 0.7~M NaCl (black dotted lines), water-glycerol mixtures containing 0.9~M NaCl (blue solid lines), water-DMSO mixtures containing 0.7~M NaCl (green solid lines). The numbers indicate the volume fractions (in $\mathrm{vol.}\%$) of glycerol and DMSO.}
\end{figure}
%--------------------------------------------------------------------------

\begin{figure}[t!]
\includegraphics{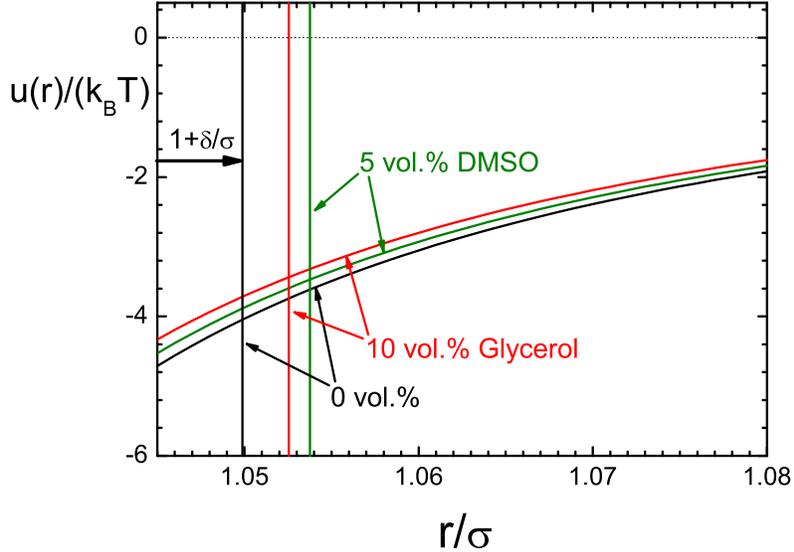}
\caption{\label{plot_potential_cs07}DLVO pair potential, $u(r)$, for NaCl concentration $c_{s}=0.7\,\mathrm{M}$, protein volume fraction $\phi=0.002$, and temperature $T=20^{\circ}\mathrm{C}$, for an aqueous solution (black line), a water-glycerol mixture ($10\,\mathrm{vol.}\%$, red line), and a water-DMSO mixture ($5\,\mathrm{vol.}\%$, green line). The vertical lines indicate the cut-off distances $\delta$ of the van der Waals attraction.}
\end{figure}

We have determined the zero- and non-zero frequency parts of the Hamaker constant (Eq.~(\ref{Hamaker_constant})) based on the above values of the refractive indices and static dielectric constants. 
The non-zero frequency part, $A_{\nu>0}$, is about ten times larger than the zero-frequency part, $A_{\nu=0}$, and hence dominates $A$ (Fig.~\ref{plot_Hamaker_constant}). 
Both parts decrease with increasing temperature as well as increasing salt, glycerol and DMSO content. 
Glycerol and DMSO cause about the same decrease of $A_{\nu>0}$, reflecting a similar effect on $n_{\mathrm{s}}$ (Eq.~(\ref{Hamaker_constant}) and Fig.~\ref{refractive_index_solvent}). 
In contrast, we find a less significant decrease of $A_{\nu=0}$ upon addition of DMSO than of glycerol, due to the weak effect of DMSO on $\varepsilon_{\mathrm{s}}$ (Fig.~\ref{dielectric_constant}b). 
Added salt (NaCl) causes a pronounced decrease of $A_{\nu=0}$ since salt significantly lowers $\varepsilon_{\mathrm{s}}$ ({Fig.~\ref{dielectric_constant}}c). 
For our sample compositions, i.e.~solvent mixtures with added salt, the temperature dependence of $\varepsilon_{s}$ is not available. 
However, since $A_{\nu =0}$ is much smaller than $A_{\nu > 0}$, we neglect the effect of added salt and consider only the effect of glycerol and DMSO. 
Nevertheless, in our calculations we take into account the effect of added salt on $n_{\mathrm{s}}$ and thus $A_{\nu>0}$ (Fig.~\ref{plot_Hamaker_constant}b), which represents the dominating contribution to the Hamaker constant.  
Glycerol and DMSO not only similarly affect $A_{\nu > 0}$ and thus the Hamaker constant $A$, but also the total DLVO pair potential, $u(r)$, (see {Fig.~\ref{plot_potential_cs07}}).

%--------------------------------------------------------------------------
%--------------------------------------------------------------------------

\bigskip

\paragraph{Cut-off length $\delta$.}
There is one potential parameter left to be determined, namely the cut-off length, $\delta$. 
We choose it such that it reproduces the second virial coefficient, $B_2(T)$, which we have experimentally determined by static light scattering.

From a theoretical point of view, the second virial coefficient, $B_2(T)$, is introduced by expanding the reduced free energy density, $f(T,\phi)$, in a power series of the particle volume fraction, $\phi$, according to
\begin{equation}\label{virial_expansion}
  f(T,\phi)=f^{\mathrm{id}}_{\mathrm{0}}(\phi)+ 4 b_{2}(T) \phi^2+\mathcal{O}(\phi^3)\,\mbox{,}
\end{equation}
\noindent which involves the ideal gas part of the free energy density, $f^{\mathrm{id}}_{\mathrm{0}}$, and its leading-order correction, the normalized or reduced second virial coefficient
\begin{equation}\label{b2_definition} 
  b_{2}(T)=\frac{B_{2}(T)}{B_{2}^{(0)}}\,\mbox{.}
\end{equation}
\noindent The latter is equal to the ratio of the second virial coefficient,
\begin{equation}\label{B2_equation}
  B_{2}=-\frac{1}{2}\int\! d{\bf r}\,{\left( \exp{\left[ -\beta u(r)\right]} - 1 \right)}\,\mbox{,}
\end{equation}
to the second virial coefficient of a hard-sphere system, $B_{2}^{(0)}=2\pi ( \sigma + \delta )^3 / 3$.
For the DLVO pair potential, $u(r)$, (in {Eq.~(\ref{total_pair_potential})}) the normalized second virial coefficient, $b_{2}$, can be written as
\begin{equation}\label{b2_decomposed}
  b_{2} = 1 - \frac{3}{(\sigma + \delta)^3}\int_{\sigma+\delta}^{\infty}\! dr\, r^2\,\,{\left( \exp{\left[ -\beta ( u_{\mathrm{el}}(r) + u_{\mathrm{vdW}}(r) ) \right]} - 1 \right)}\,\mbox{.} 
\end{equation}
\noindent The second virial coefficient characterizes the thermodynamic strength of the pair interactions. 
Positive values of $b_{2}$ indicate that repulsive interactions dominate, while negative values correspond to dominating attraction.

%--------------------------------------------------------------------------
\begin{figure}[t!]
\includegraphics{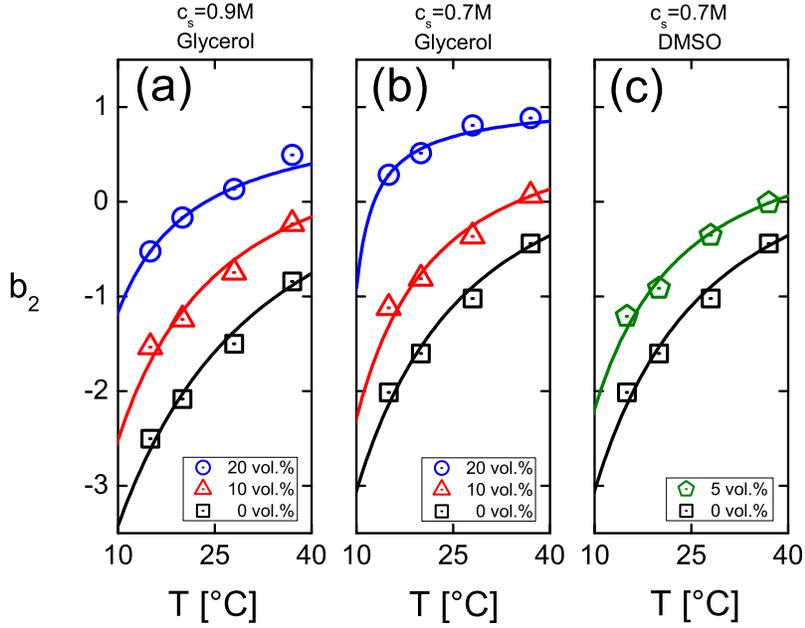}
\caption{\label{b2_glycerol_DMSO} Temperature dependence of the normalized second virial coefficient, $b_{2}(T)$, for different added NaCl concentrations, $c_{\mathrm{s}}$, as well as glycerol and DMSO contents (as indicated), experimentally determined by static light scattering. The solid lines represent fits assuming a linear temperature dependence of the cut-off length $\delta$.}
\end{figure}
%--------------------------------------------------------------------------

Using SLS, we determined the temperature dependence of $b_{2}(T)$ in the range $10^{\circ}\mathrm{C} < T < 40^{\circ}\mathrm{C}$ for all relevant solution compositions, i.e.~salt (NaCl) concentrations, $c_{\mathrm{s}}$, as well as glycerol and DMSO volume fractions ({Fig.~\ref{b2_glycerol_DMSO}}). 
Outside this temperature range, reliable SLS experiments cannot be performed.
With increasing temperature, as expected $b_{2}(T)$ and thus the repulsion increases, which reflects the increasing electrostatic (Fig.~\ref{plot_contact_value_Coulomb_repulsion}) and decreasing van der Waals interactions (Fig.~\ref{plot_Hamaker_constant}). 
This leads also to an increase of $b_{2}(T)$ with increasing glycerol and DMSO concentrations, in agreement with previous observations \cite{kulkarni:2002,liu:2004,sedgwick:2007}, with DMSO having a significantly more pronounced effect.
Furthermore, $b_{2}(T)$ increases on lowering $c_{\mathrm{s}}$, due to the increased electrostatic repulsion ({Figs.~\ref{b2_glycerol_DMSO}}a and b). 
The sample with $c_{\mathrm{s}}=0.7\,\mathrm{M}$ and $20\,\mathrm{vol.}\%$ glycerol even shows a positive $b_{2}(T)$ in the investigated temperature range $10^{\circ}\mathrm{C} < T < 40^{\circ}\mathrm{C}$, indicating that the lysozyme pair potential is predominantly repulsive at high glycerol concentrations.

\begin{figure}[t!]
\includegraphics{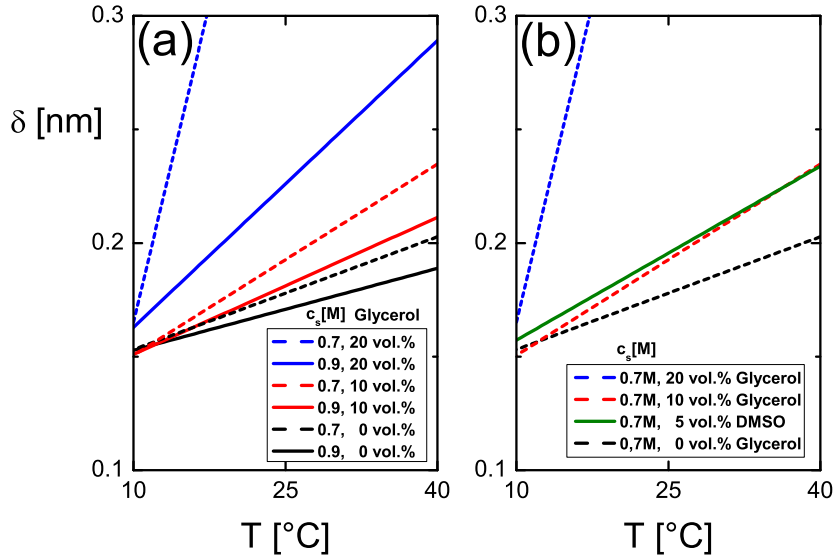}
\caption{\label{plot_delta}Temperature dependence of the cut-off length $\delta(T)$ as obtained from fits of DLVO-based predictions to the measured $b_{2}(T)$. (a) Different added NaCl concentrations, $c_{\mathrm{s}}$, and glycerol contents and (b) same NaCl concentration $c_{\mathrm{s}}=0.7\,\mathrm{M}$, but different glycerol and DMSO contents (as indicated).}
\end{figure}

The experimental $b_{2}(T)$ is fitted by the DLVO potential prediction ({Eq.~(\ref{b2_decomposed})}), with the cut-off length $\delta$ being the only adjustable parameter. 
A constant $\delta$ does not reproduce the experimental data satisfactorily. 
We thus assume a linear temperature dependence of the cut-off length, $\delta(T)=\delta_{0}+\delta_{1}T$, and fit the coefficients $\delta_{0}$ and $\delta_{1}$ to obtain optimal agreement between experimental and theoretical $b_{2}(T)$ in the temperature range $10^{\circ}\mathrm{C} < T < 40^{\circ}\mathrm{C}$ ({Fig.~\ref{b2_glycerol_DMSO}}). 
The fitting is done by applying the Levenberg-Marquardt method \cite{press:1992}, and the improper integrals appearing in the expression for the second virial coefficient are evaluated using the Chebyshev quadrature. 
For all considered compositions and the temperature range of the light scattering experiments, $\delta(T)$ increases with increasing $T$ ({Fig.~\ref{plot_delta}}). 
Its values are in the range $0.15\,\mathrm{nm} \le \delta(T) \le 0.3\,\mathrm{nm}$, except for the solution with $0.7\,\mathrm{M}$ NaCl and  $20\,\mathrm{vol.}\%$ glycerol for which unphysically large $\delta(T)$ are obtained. 
This range of values agrees with previously published values, which are in the range of $0.1\,\mathrm{nm} \le \delta(T) \le 0.3\,\mathrm{nm}$ \cite{muschol:1995,kuehner:1997,velev:1998,tardieu:1999,egelhaaf:2004}. 
The large values for $\delta(T)$ obtained for $0.7\,\mathrm{M}$ NaCl and $20\,\mathrm{vol.}\%$ glycerol are due to the positive-valued experimental $b_{2}(T)$, while the DLVO model predicts a predominantly attractive potential and thus a negative $b_{2}(T)$.
This forces $\delta(T)$ to  rapidly grow with $T$ to reduce the attractive van der Waals contribution. 
The effect of the van der Waals interaction on $b_{2}(T)$ is controlled by the Hamaker constant $A$ and the cut-off length $\delta$. 
A change in one of the two hence can be compensated (or enhanced) by a corresponding change in the other parameter.

%--------------------------------------------------------------------------
%--------------------------------------------------------------------------

The cut-off length, $\delta$, accounts for molecular effects on a coarse-grained level, such as the solvation shell, which might contain both water and additive molecules (see Sec.~\ref{sec:intro}), or the condensation of counterions on the protein surface, {i.e.} in the Stern layer. 
Its increase with increasing temperature and glycerol content is consistent with previous findings \cite{sinibaldi:2007, esposito:2009, vagenende:2009}. 
As mentioned above, glycerol affects $b_{2}(T)$ less than DMSO, i.e.~the addition of $10\,\mathrm{vol.}\%$ glycerol has about the same effect as $5\,\mathrm{vol.}\%$ DMSO (Figs.~\ref{b2_glycerol_DMSO}b and c), although the optical properties of the two solvents suggest a similar effect on $A_{\nu > 0}$ and thus $A$ (Fig.~\ref{plot_Hamaker_constant}) and, due to the small effect of glycerol and DMSO on the electrostatic interactions (Fig.~\ref{plot_contact_value_Coulomb_repulsion}), also on the DLVO potential (Fig.~\ref{plot_potential_cs07}). 
This results in a larger cut-off length $\delta(T)$ for DMSO for a given concentration of glycerol or DMSO ({Fig.~\ref{plot_delta}}), and suggests that glycerol and DMSO interact differently with lysozyme. 
Compared to glycerol, little is known about the DMSO-lysozyme interactions, apart from a preferential hydration of lysozyme in water-DMSO mixtures indicating a weak or no binding of DMSO to lysozyme \cite{arakawa:2007, kamiyama:2009}. 
This suggests that DMSO mainly influences the dielectric properties of the solution, and is consistent with the observed weaker temperature dependence of the cut-off length $\delta(T)$ in the case of DMSO.

%%%%% Phase Behavior

\subsection{Phase Behavior}
\label{sec:phase}

We have now determined all values of the parameters determining the DLVO model. 
We can thus predict the phase behavior without any free parameters and compare it to our experimental observations.

%%%%% TPT

\subsubsection{\label{TPT}Phase Behavior and Free Energy by Thermodynamic Perturbation Theory}

Our prediction of the phase behavior of lysozyme is based on expressions for the Helmholtz free energy which are obtained by the second-order thermodynamic perturbation theory (TPT) of Barker and Henderson \cite{barker:1976}.
Knowledge of the free energy of the fluid and solid phases allows us to deduce all equilibrium thermodynamic properties, such as the fluid-solid and gas-liquid coexistence curves including the critical point. 
The Barker-Henderson theory is known to yield quite accurate results \cite{hagen:1994,cochran:2004}. 
It is therefore widely used to calculate the phase diagram of protein solutions \cite{tavares:2004,bostrom:2006,lima:2007,gogelein:2008,dorsaz:2009}.

Within TPT, the free energy, $F$, is expanded in powers of the perturbation potential $u_{\mathrm{p}}= u - u_{\mathrm{0}}$, where $u_{0}$ is the pair potential of a reference system and $u$ the total protein-protein pair potential (see {Eq.~(\ref{total_pair_potential})}). 
The reference system is an effective hard-sphere system with diameter $\sigma + \delta$, which takes into account the cut-off length $\delta$ in {Eq.~(\ref{eq_vdW_interaction})}.
Instead of the free energy, $F(N,V,T)$, of a system with $N$ proteins in a volume $V$, we use the dimensionless free energy density $f=\beta F v_{\mathrm{0}}/V$, where $v_{\mathrm{0}}=\pi\sigma^3/6$ is the particle volume.
Expanding the free energy density, $f$, up to second order in $u_{\mathrm{p}}$ gives
\begin{eqnarray}\label{reduced_free_energy}
  f(\phi,T) = f_{0}(\phi) & + & \frac{12}{{(\sigma+\delta)}^3}\,{\phi'}^2\int_{\mathrm{\sigma+\delta}}^{\mathrm{\infty}}\! dr \, r^2 g_{0}(r)\,\beta u_{\mathrm{p}}(r) \nonumber\\
  & - & \,\frac{6}{{(\sigma+\delta)}^3}\,{\phi'}^2 {\left(\frac{\partial \phi}{\partial {\Pi}_{0}}\right)}_{\mathrm{T}}\int_{\mathrm{\sigma+\delta}}^{\mathrm{\infty}}\! dr\, r^2\, g_{\mathrm{0}}(r) {\left[ \beta u_{\mathrm{p}}(r)\right]}^2 \,\,\mbox{,}  
\end{eqnarray}
\noindent where $\phi'=\phi(1+\delta/\sigma)^3$ is the effective sphere volume fraction, and $\chi_{\mathrm{T}}/(\beta v_{0})=1/\phi {(\partial\phi/\partial{\Pi}_{0})}_{\mathrm{T}}$ the isothermal compressibility of the reference system with $\Pi_{0}$ its dimensionless osmotic pressure. 
Furthermore, $g_{0}(r)$ is the radial distribution function of hard spheres in the fluid phase, and the orientationally-averaged pair distribution function of hard spheres in the solid phase. 
For the radial distribution function in the fluid phase, we use the Verlet-Weis corrected \cite{verlet:1972} Percus-Yevick solution \cite{wertheim:1963,throop:1965}, and for the orientationally-averaged pair distribution function the expression by Kincaid \cite{kincaid:1977} for a face-centered cubic (fcc) crystal phase. 
The free energy density, $f_{\mathrm{0}}(\phi)$, of the effective hard-sphere reference system is separately defined in the fluid and solid branches.
In the fluid phase, $f_{\mathrm{0}}(\phi)$ consists of two parts: First, the ideal gas part,
\begin{equation}\label{free_energy_ideal_gas}
  \alpha\,f_{0}^{\mathrm{id}}(\phi) = \phi' \left[ \ln( \phi' \Lambda^3 / v_{0}) - 1 \right]\,\mbox{,}
\end{equation}
\noindent where $\Lambda=h/\sqrt{2\pi m_{\mathrm{p}} k_{\mathrm{B}}T}$ is the thermal wavelength,  $m_{\mathrm{p}}$ the protein mass and  $\alpha=\phi' /\phi$. 
Second, the interaction part which is approximated by the accurate Carnahan-Starling equation of state \cite{carnahan:1969}
\begin{equation}
  \alpha\,f_{\mathrm{0}}^{\mathrm{CS}}(\phi) = \frac{4{\phi'}^2 - 3{\phi'}^3}{(1-{\phi'})^2}\,\mbox{.}
\end{equation}
\noindent In the solid phase, $f_{\mathrm{0}}(\phi)$ is described by Wood's equation of state \cite{wood:1952} on assuming a fcc crystalline lattice
\begin{equation}\label{f_HS_Wood}
  \alpha\,f_{\mathrm{0}}^{^{\mathrm{solid}}}(\phi) = 2.1306\,\phi' + 3\,\phi'\ln{\left( \frac{\phi'}{1-{\phi'}/{\phi}_{\mathrm{cp}}}\right)} + {\phi'} \ln{\left(\frac{\Lambda^3}{v_{\mathrm{0}}}\right)}\,\mbox{,}
\end{equation}
\noindent where  $\phi_{\mathrm{cp}}=\pi\sqrt{2}/6$ is the fcc volume fraction for closed packing. 
To evaluate the free energy density, $f(\phi,T)$, in the solid phase, the integrals in {Eq.~(\ref{reduced_free_energy})} are divided into intervals centered around the crystal lattice sites, which is necessary to sufficiently resolve the discreteness of $g_{0}(r)$ that develops with increasing $\phi$.

%%%%% Phase Coexistence

The coexistence of gas (g) and liquid (l) phases requires an identical temperature $T$, osmotic pressure $\Pi$ and chemical potential $\mu$ in both phases. 
While the temperatures of two phases in contact are always identical under equilibrium conditions, the latter two conditions determine the two volume fractions, namely that of the coexisting gas, $\phi_{\mathrm{g}}$, and liquid phase, $\phi_{\mathrm{l}}$:
\begin{eqnarray}\label{equal_pressure_g_l}
  \Pi_{\mathrm{g}}(T,\phi_{\mathrm{g}})=\Pi_{\mathrm{l}}(T,\phi_{\mathrm{l}})\quad & \mbox{with}\quad &\Pi(T,\phi)=\phi^2\left(\frac{\partial(f(T,\phi)\,/\,\phi)}{\partial \phi}\right)_{\mathrm{T}} \\
\label{equal_chem_pot_g_l}
  \mu_{\mathrm{g}}(T,\phi_{\mathrm{g}})=\mu_{\mathrm{l}}(T,\phi_{\mathrm{l}})\quad & \mbox{with}\quad &\beta\mu (T,\phi )=\left(\frac{\partial f(T,\phi)}{\partial \phi}\right)_{\mathrm{T}}\,\mbox{.}
\end{eqnarray}
Similarly, the coexistence of fluid (f) and solid (s) phases is determined by the two conditions    
\begin{eqnarray}\label{equal_pressure}
  \Pi_{\mathrm{f}}(T,\phi_{\mathrm{f}})=\Pi_{\mathrm{s}}(T,\phi_{\mathrm{s}}) \\
\label{equal_chem_pot}
  \mu_{\mathrm{f}}(T,\phi_{\mathrm{f}})=\mu_{\mathrm{s}}(T,\phi_{\mathrm{s}})\,\mbox{,}
\end{eqnarray}
where $\phi_{f}$ is the volume fraction of the fluid, and $\phi_{s}$ the one of the coexisting solid phase. 
To compute the phase coexistence curves, the Newton-Raphson method with line search is applied \cite{press:1992}.

The critical point terminates the gas-liquid coexistence curve. 
It is determined by the vanishing second and third derivatives of the free energy density, {i.e.}, by
\begin{equation}
  \frac{{\partial}^2f(T_{c},\phi_{c})}{{\partial}^2\phi}=0 \quad \mbox{and} \quad  \frac{{\partial}^3 f(T_{c},\phi_{c})}{{\partial}^3\phi}=0\,\mbox{.} 
\end{equation}  
The higher-order derivatives of the free energy are calculated using Ridder's implementation of Neville's algorithm \cite{press:1992}.

%%%%% Test of our TPT Implementation

Before comparing our theoretical and experimental findings, we test the accuracy of our TPT implementation. 
No simulated phase diagrams are available for the DLVO model with our parameter values. 
However, Monte Carlo simulations \cite{hagen:1994,shukla:2000} have been used to determine the phase diagram of hard spheres interacting via an attractive Yukawa potential
\begin{equation}\label{attr_Yukawa_potential}
  u(r) = \left\{
                \begin{array}{ll}
                  \infty\,\mbox{,} & r < \sigma \\
                  -\xi\,\frac{\exp[-z (r-\sigma )]}{r/\sigma}\,\mbox{,} & r \geq \sigma\,\mbox{,}
                \end{array}
         \right. 
\end{equation} 
\noindent where $\xi>0$ is the (contact) strength, and $z^{-1}$ the range of the attraction.

%--------------------------------------------------------------------------
\begin{figure}[t!]
\includegraphics[width=0.9\textwidth]{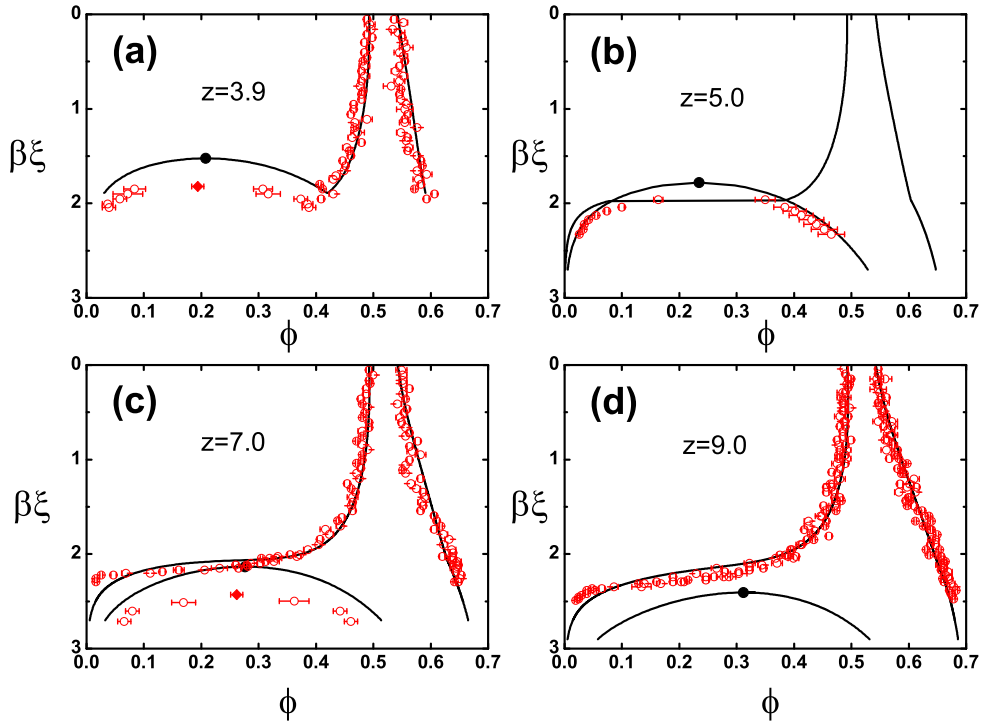}
\caption{\label{pd_attr_Yukawa}Phase diagrams with fluid-solid and gas-liquid coexistence curves of attractive Yukawa potential systems with different ranges of attraction, $z^{-1}$, as indicated. Coexistence curves with critical points calculated using TPT are represented as solid lines with black filled circles, and the data from Monte Carlo simulations are represented as red open circles \cite{hagen:1994,shukla:2000}. The coexistence curves are shown in the normalized density ($\rho \sigma^3$) -- normalized potential strength ($\beta \xi$) -- plane. Note that the ($\beta\xi$) -- axes run downwards. Furthermore, no simulation data for the fluid-solid and gas-liquid coexistence curves for $z=5.0$ and $z=9.0$, respectively, were provided in \cite{hagen:1994,shukla:2000}.}
\end{figure}
%--------------------------------------------------------------------------

Using TPT, we calculate the phase diagram of this system.
The calculated fluid-solid coexistence curves agree with the simulated ones for all investigated values of $z^{-1}$ (see {Fig.~\ref{pd_attr_Yukawa}}).
Also the widths of the gas-liquid coexistence curves are well reproduced. 
The calculated gas-liquid binodals, however, are located slightly above the simulation data. 
We suppose that the binodals are shifted because the mean-field type TPT does not account for critical density fluctuations. 
Critical fluctuations allow a system to probe density inhomogeneities and thus tend to stabilize the homogeneous supercritical fluid phase against phase separation, therefore shifting the critical point to lower temperatures. 
A renormalization-group correction scheme has been derived \cite{lue:1998} which, according to previous calculations \cite{fu:2003}, is expected to flatten the coexistence curves around the critical point.

%%%%% Experimentally Determined Phase Behavior

\subsubsection{\label{results_and_discussion}Experimentally Determined Phase Behavior}

\begin{figure}[t!]
\includegraphics{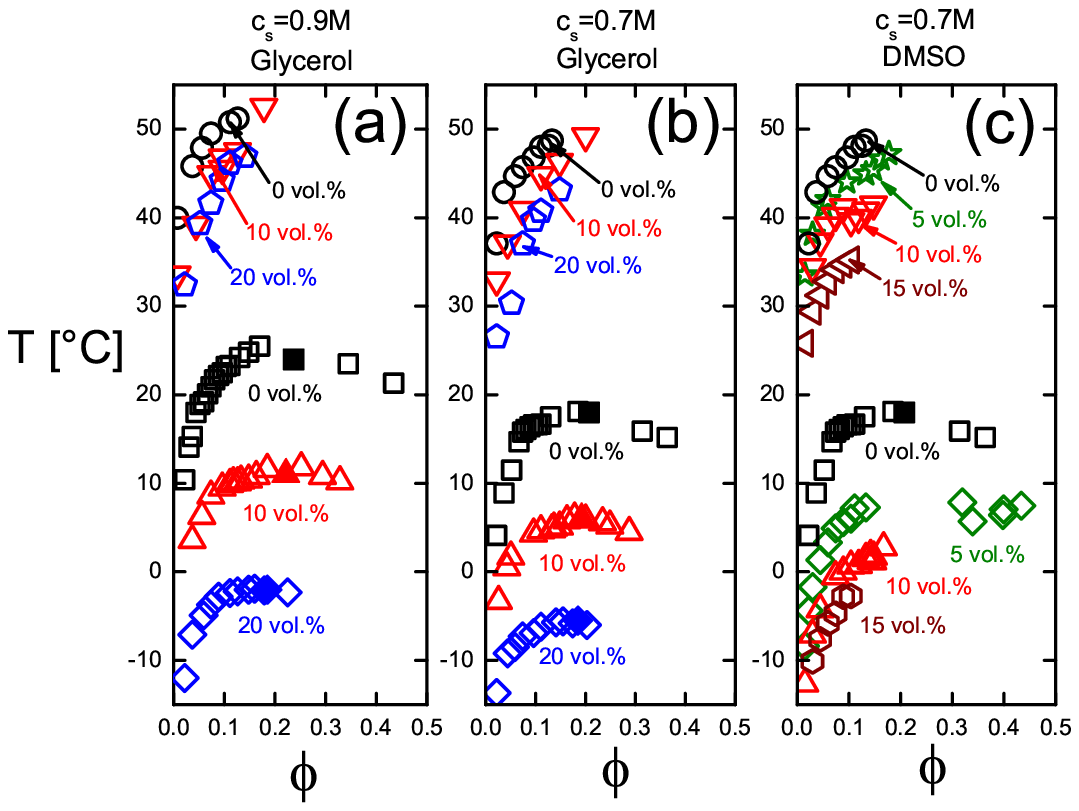}
\caption{\label{plot_pd_glycerol_DMSO}Experimental phase diagram of lysozyme solutions with different added NaCl concentrations, $c_{\mathrm{s}}$, as well as different glycerol and DMSO contents (as indicated), in dependence on the volume fraction, $\phi$, and temperature, $T$. The open circles, inverted triangles, pentagons, stars and rotated triangles indicate the fluid-solid coexistence curves. The open squares, triangles, diamonds, hexagons and octagons indicate the gas-liquid coexistence curves. The filled squares, triangles and diamonds indicate the critical points.}
\end{figure}

We experimentally determined the fluid-solid and gas-liquid coexistence curves for different contents of added salt (NaCl), glycerol and DMSO ({Fig.~\ref{plot_pd_glycerol_DMSO}}). 
For all compositions explored, the gas-liquid binodals lie within the fluid-solid coexistence regions and are thus metastable. 
This is typical for short-ranged attractive pair potentials \cite{broide:1996,asherie:1996,rosenbaum:1996b,muschol:1997}.
The binodal widens rapidly upon cooling at volume fractions $\phi$ below the critical volume fraction $\phi_{\mathrm{c}}$, while the data for $\phi > \phi_{\mathrm{c}}$ suggest almost flat binodals at volume fractions larger than the critical point value.
Similar shapes of lysozyme binodals were observed previously \cite{muschol:1997, rosenbaum:1996}.

%--------------------------------------------------------------------------
%--------------------------------------------------------------------------

The effect of salt on the phase behavior was studied for $0.7$ and $0.9\,\mathrm{M}$ NaCl ({Figs.~\ref{plot_pd_glycerol_DMSO}}a and b, respectively). 
With increasing salt concentration, $c_{\mathrm{s}}$, the fluid-solid and gas-liquid coexistence curves move to higher temperatures, and the width of the gas-liquid binodal increases. 
These findings are consistent with the electrostatic screening by added salt. 
Upon increasing $c_{\mathrm{s}}$, the screening length, ${\kappa}^{-1}$, decreases. 
Thus the electrostatic repulsion decreases and attraction becomes more important. 
This favors crystals compared to the fluid phase and also the (metastable) gas-liquid phase separation compared to a homogeneous fluid. 
The fluid-solid as well as the gas-liquid coexistence curves hence shift to higher temperatures, and the gas-liquid coexistence curve becomes broader, as experimentally observed.

Upon addition of glycerol and DMSO, both coexistence curves shift to lower temperatures (see {Fig.~\ref{plot_pd_glycerol_DMSO}}).
The shift of the binodals is more pronounced, which widens the gap between the two coexistence curves.
This trend has been observed before in other systems and is characteristic for a decreasing range of attraction \cite{broide:1996,hagen:1994,liu:2005} (Fig.~\ref{plot_potential_cs07}).
The critical point is not only shifting to lower temperatures, but also to slightly lower volume fractions. 
Considering the quantitative effect on both curves, we find that the addition of $10\,\mathrm{vol.}\%$ glycerol and $5\,\mathrm{vol.}\%$ DMSO induce a similar shift. 
The higher efficiency of DMSO is consistent with its more pronounced effect on $b_{2}(T)$ (Fig.~\ref{b2_glycerol_DMSO}).

%--------------------------------------------------------------------------
\begin{figure}[t!]
\includegraphics[height=13cm]{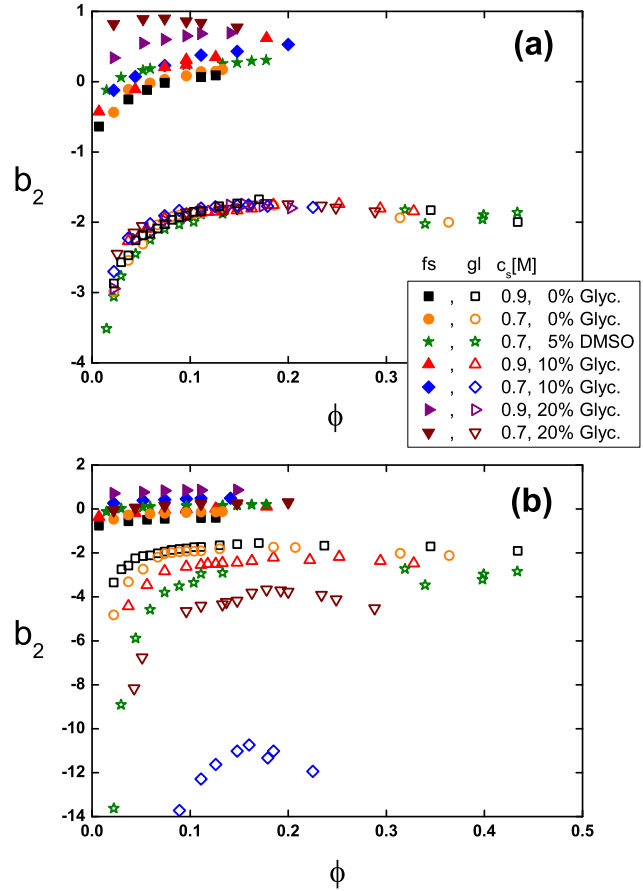}
\caption{\label{b2_phi}Experimental phase diagram of lysozyme solutions with different added NaCl concentrations, $c_{\mathrm{s}}$, as well as different glycerol and DMSO contents (as indicated), as a function of volume fraction, $\phi$, and reduced second virial coefficient, $b_{2}$. The second virial coefficient has been (a) experimentally determined and (b) calculated based on the DLVO model. The filled symbols indicate the fluid-solid (fs) coexistence curves, the open symbols the gas-liquid (gl) coexistence curves.}
\end{figure}
%--------------------------------------------------------------------------

We have investigated the relation between the second virial coefficient, $b_{2}(T)$, and the location of the fluid-solid and especially the gas-liquid coexistence curves. 
For this purpose, we parameterized the temperature dependencies of the experimentally-determined second virial coefficients (Fig.~\ref{b2_glycerol_DMSO}) by the phenomenological relation $b_{2}(T)=\alpha_0+\alpha_1\,T +\alpha_2\,T^2$, where $\alpha_1$ and $\alpha_2$ depend on the composition of the solution. 
Using this relation, the binodals are plotted in dependence of $b_{2}$ and $\phi$ in Fig.~\ref{b2_phi}a. 
This representation leads to a collapse of all binodals onto a single curve, as it has been observed previously \cite{george:1994, rosenbaum:1996, poon:2000, farnum:1999}. 
This indicates that in our system the binodals are controlled by $b_2$, i.e.~by a single integral parameter characterizing the potential, and not by the details of the potential, such as its depth and range.
By contrast, this scaling of $b_{2}$ is not observed for the fluid-solid transition.
Nevertheless, this transition occurs for $b_2$ values consistent with previous reports \cite{poon:1997}.

If instead the $b_2$ values calculated based on the DLVO model are used, the fluid-solid and gas-liquid coexistence curves do not scale (see Fig.~\ref{b2_phi}b) and the critical points are different.
This indicates that the scaling very sensitively depends on the temperature dependence of $b_{2}(T)$ and thus on the model. 
Small errors might be introduced by the simplifications implicit in the DLVO model. 
However, the main effect is due to the extrapolation of $b_{2}(T)$ beyond the available temperature range $10^{\circ}\mathrm{C} < T < 40^{\circ}\mathrm{C}$ (see Fig.~\ref{b2_glycerol_DMSO}), which indicates the need for a very reliable and robust parameterization of the temperature dependence of the cut-off length $\delta(T)$. 
To avoid spurious effects, in the following we scale the temperature and concentration axes by the critical temperatures and concentrations.
This axis scaling also reduces artifacts due to the slight shift in the binodals caused by the mean-field nature of TPT (Fig.~\ref{pd_attr_Yukawa}).

%%%%% Comparison of Calculated and Experimentally Determined Phase Behavior

\subsubsection{Comparison of Calculated and Experimentally Determined Phase Behavior}
\label{sec:comp}

%--------------------------------------------------------------------------
\begin{figure}[t!]
\includegraphics[height=13cm]{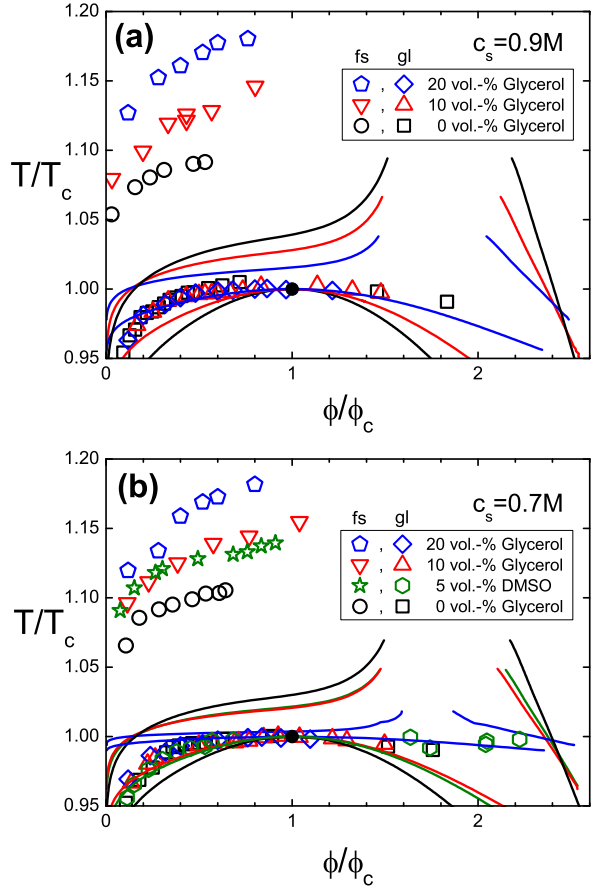}
\caption{\label{pd_rescaled} Experimental and predicted phase diagram of lysozyme solutions with different added NaCl concentrations, $c_{\mathrm{s}}$, as well as different glycerol and DMSO contents (as indicated), as a function of reduced volume fraction, $\phi/\phi_{\mathrm{c}}$, and reduced temperature, $T/T_{\mathrm{c}}$, where ($\phi_{\mathrm{c}}$,$T_{\mathrm{c}}$) defines the critical point. Solid lines represent the TPT calculations based on the DLVO model, open symbols the experimental data for the stable fluid-solid (fs) and metastable gas-liquid (gl) coexistence curves. Filled symbols mark the critical point.}
\end{figure}
%--------------------------------------------------------------------------

The experimental phase diagrams are now compared to our TPT predictions based on the DLVO model.
The axes are scaled by the critical temperature, $T_{\mathrm{c}}$, and volume fraction, $\phi_{\mathrm{c}}$, which matches all theoretical and experimental critical points (see {Fig.~\ref{pd_rescaled}}).
(Note that in this representation the fluid-solid transition shifts to higher $T/T_{\mathrm{c}}$, in contrast to the representation using the absolute temperature scale (Fig.~\ref{plot_pd_glycerol_DMSO}).
This representation reveals an increase of the gap between the fluid-solid and the gas-liquid coexistence curves upon addition of glycerol and DMSO for the experimental data and, less pronounced, in the range $\phi/\phi_{\mathrm{c}} < 0.2$ for the theoretical predictions. Note that in the main part of the phase diagram the theoretically predicted gap actually decreases.
In view of the successful test of the TPT approach (Fig.~\ref{pd_attr_Yukawa}), we attribute this discrepancy to the inherent approximations of the DLVO potential and in particular to the model for the cut-off length $\delta$.

The width of the gas-liquid coexistence curve is experimentally found to remain practically unchanged in this representation when glycerol or DMSO is added, while theory predicts a slight increase of the width.
The theoretically and experimentally observed widths quantitatively agree at the largest glycerol content of $20\,\mathrm{vol.}\%$ ({Fig.~\ref{pd_rescaled}}).
In particular, our calculations do not reproduce the flat binodal region around the critical point, which we observe experimentally.
As discussed above (see Fig.~\ref{pd_attr_Yukawa}), this is attributed to the mean-field character of the theory.
This is furthermore consistent with previous experiments \cite{muschol:1997} and simulations \cite{pellicane:2003}.
In particular, similar gas-liquid coexistence curves have been theoretically observed for colloidal dispersions with competing repulsive and attractive interactions, and have been attributed to critical fluctuations which become significant in an enlarged region of the phase diagram \cite{archer:2007}.

%%%%% Conclusions %%%%%
 
\section{\label{conclusion}Conclusions}

In a combined theoretical and experimental effort, we investigated the phase behavior of aqueous lysozyme solutions in the presence of additives, namely glycerol and DMSO.
We in particular investigated the fluid-solid and gas-liquid coexistence curves.
Upon the addition of glycerol or DMSO, both curves shift to lower temperatures and, in addition, the gap between these two curves increases, consistent with a decreasing range of attraction upon increasing the content of additives.

Our theoretical calculations are based on a DLVO interaction potential.
This models the proteins as hard spheres with an isotropic electrostatic repulsion, which is caused, {e.g.}, by uniformly-distributed charges, and an isotropic van der Waals attraction characterized by the (effective) Hamaker constant and a cut-off length.
The Hamaker constant is calculated based on the experimentally determined, or in the literature available, macroscopic optical and dielectric properties of the solvent mixtures and the protein, namely their indices of refraction and static dielectric constants.  
The cut-off length $\delta$ was adjusted to reproduce the second virial coefficient, which was experimentally determined by static light scattering.
We find cut-off lengths in the range $0.15\,\mathrm{nm} \le \delta(T) \le 0.3\,\mathrm{nm}$, with the value increasing with increasing temperature and glycerol content and, even more pronounced, with increasing DMSO content, consistent with previous studies \cite{muschol:1995,kuehner:1997,velev:1998,tardieu:1999,egelhaaf:2004}.
The cut-off length subsumes molecular effects, such as a hydration shell or condensed counterions on the protein surface.
Our experiments thus indicate that glycerol and DMSO interact differently with lysozyme.
DMSO  seems to affect the lysozyme interactions by mainly lowering the solvent dielectric constant, while glycerol likely changes the hydrophobicity of the lysozyme surface.
This is in accordance with recent computer simulation results \cite{vagenende:2009}.

Having determined the values of all parameters, we calculated the interaction potential for different temperatures and compositions, added salt (NaCl) concentrations as well as different glycerol and DMSO contents.
We found that glycerol and DMSO increase the repulsive interaction, as quantified by the second virial coefficient $b_2$.
For a sufficiently high content of additives, a predominantly repulsive protein interaction is observed.
The increase of the repulsion is significantly stronger for DMSO, although the optical and dielectric properties of glycerol and DMSO are very similar.
We hence attribute this difference to the specific molecular interactions of the additives with lysozyme, characterized by the cut-off length $\delta(T)$.

Based on the DLVO model and its parameters, we predict the phase behavior without any free parameters.
Our calculations are based on the thermodynamic perturbation theory (TPT), which was successfully tested and found to be numerically fast.
The theoretical predictions were then compared to our experimental observations, for the fluid-solid and the gas-fluid coexistence curves.
Particular attention was paid to the effect of additives on these curves, and the size of the gap between them.
We find that the general trends in the phase behavior, and especially the effects of the additives, are also reproduced by the DLVO model, although not on a quantitative level.

The strength of the DLVO model lies in its simplicity, for it allows to incorporate the effect of solvent mixtures by taking into account their macroscopic optical and dielectric properties.   
More complex models are available, but might need to be adapted to proteins and solvent mixtures or additives. 
For example, theoretical calculations based on anisotropic attractive interactions successfully describe both the width of the binodal and the gap between the binodal, and the fluid-solid transition for aqueous protein solutions without additives \cite{lomakin:1999,gogelein:2008,gogelein:2011}.    
Anisotropic attractive interactions are indeed induced by hydrophobic patches on the lysozyme surface, and have been taken into account in a patchy sphere model by Curtis {\it et al.}~\cite{curtis:2002}.
Anisotropic attractions as well as repulsions, as due {e.g.} to an anisotropic charge distribution or van der Waals interaction, are not taken into account in our DLVO model. 
Our model furthermore neglects non-electrostatic contributions to the lysozyme pair potential that originate from dispersion forces between the micro-ions and proteins \cite{bostrom:2006,lima:2007}. 
The range and strength of such non-electrostatic and hydrophobic interactions has not been determined experimentally, and hence cannot be taken into account unambiguously, {e.g.}, through an effective potential of mean force.
Furthermore, hydrophobic interactions, as well as non-rigid and non-spherical shapes, specific surface properties, ion-specific interactions, hydration, and other effects need to be explored in future studies.

%%%%% Acknowledgements %%%%%

\section{Acknowledgements}

We thank the International Helmholtz Research School of Biophysics and Soft Matter (IHRS BioSoft) and the Deutsche Forschungsgemeinschaft (DFG) for financial support.

%%%%% References %%%%%


\begin{thebibliography}{81}
\expandafter\ifx\csname natexlab\endcsname\relax\def\natexlab#1{#1}\fi
\expandafter\ifx\csname bibnamefont\endcsname\relax
  \def\bibnamefont#1{#1}\fi
\expandafter\ifx\csname bibfnamefont\endcsname\relax
  \def\bibfnamefont#1{#1}\fi
\expandafter\ifx\csname citenamefont\endcsname\relax
  \def\citenamefont#1{#1}\fi
\expandafter\ifx\csname url\endcsname\relax
  \def\url#1{\texttt{#1}}\fi
\expandafter\ifx\csname urlprefix\endcsname\relax\def\urlprefix{URL }\fi
\providecommand{\bibinfo}[2]{#2}
\providecommand{\eprint}[2][]{\url{#2}}

\bibitem[{\citenamefont{McPherson}(1999)}]{mcpherson:1999}
\bibinfo{author}{\bibfnamefont{A.}~\bibnamefont{McPherson}},
  \emph{\bibinfo{title}{Crystallization of Biological Macromolecules}}
  (\bibinfo{publisher}{Cold Spring Harbor}, \bibinfo{address}{New York},
  \bibinfo{year}{1999}), \bibinfo{edition}{1st} ed.

\bibitem[{\citenamefont{Sousa}(1995)}]{sousa:1995}
\bibinfo{author}{\bibfnamefont{R.}~\bibnamefont{Sousa}}, \bibinfo{journal}{Acta
   Crystallogr., Sect. D: Biol. Crystallogr.} \textbf{\bibinfo{volume}{D51}}, \bibinfo{pages}{271}
  (\bibinfo{year}{1995}).

\bibitem[{\citenamefont{Arakawa et~al.}(2007)\citenamefont{Arakawa, Kita, and
  Timasheff}}]{arakawa:2007}
\bibinfo{author}{\bibfnamefont{T.}~\bibnamefont{Arakawa}},
  \bibinfo{author}{\bibfnamefont{Y.}~\bibnamefont{Kita}}, \bibnamefont{and}
  \bibinfo{author}{\bibfnamefont{S.~N.} \bibnamefont{Timasheff}},
  \bibinfo{journal}{Biophys. Chem.} \textbf{\bibinfo{volume}{131}},
  \bibinfo{pages}{62} (\bibinfo{year}{2007}).

\bibitem[{\citenamefont{Kamiyama et~al.}(2009)\citenamefont{Kamiyama, Liu, and
  Kimura}}]{kamiyama:2009}
\bibinfo{author}{\bibfnamefont{T.}~\bibnamefont{Kamiyama}},
  \bibinfo{author}{\bibfnamefont{H.~L.} \bibnamefont{Liu}}, \bibnamefont{and}
  \bibinfo{author}{\bibfnamefont{T.}~\bibnamefont{Kimura}},
  \bibinfo{journal}{J. Therm. Anal. Cal.} \textbf{\bibinfo{volume}{95}},
  \bibinfo{pages}{353} (\bibinfo{year}{2009}).

\bibitem[{\citenamefont{McPherson}(1990)}]{mcpherson:1990}
\bibinfo{author}{\bibfnamefont{A.}~\bibnamefont{McPherson}},
  \bibinfo{journal}{Eur. J. Biochem.} \textbf{\bibinfo{volume}{189}},
  \bibinfo{pages}{1} (\bibinfo{year}{1990}).

\bibitem[{\citenamefont{ten Wolde and Frenkel}(1997)}]{wolde:1997}
\bibinfo{author}{\bibfnamefont{P.~R.} \bibnamefont{ten Wolde}}
  \bibnamefont{and} \bibinfo{author}{\bibfnamefont{D.}~\bibnamefont{Frenkel}},
  \bibinfo{journal}{Science} \textbf{\bibinfo{volume}{277}},
  \bibinfo{pages}{1975} (\bibinfo{year}{1997}).

\bibitem[{\citenamefont{Sedgwick et~al.}(2005)\citenamefont{Sedgwick, Kroy,
  Salonen, Robertson, Egelhaaf, and Poon}}]{sedgwick:2005}
\bibinfo{author}{\bibfnamefont{H.}~\bibnamefont{Sedgwick}},
  \bibinfo{author}{\bibfnamefont{K.}~\bibnamefont{Kroy}},
  \bibinfo{author}{\bibfnamefont{A.}~\bibnamefont{Salonen}},
  \bibinfo{author}{\bibfnamefont{M.~B.} \bibnamefont{Robertson}},
  \bibinfo{author}{\bibfnamefont{S.~U.} \bibnamefont{Egelhaaf}},
  \bibnamefont{and} \bibinfo{author}{\bibfnamefont{W.~C.~K.}
  \bibnamefont{Poon}}, \bibinfo{journal}{Eur. Phys. J. E}
  \textbf{\bibinfo{volume}{16}}, \bibinfo{pages}{77} (\bibinfo{year}{2005}).

\bibitem[{\citenamefont{Israelachvili}(1991)}]{israelachvili:1991}
\bibinfo{author}{\bibfnamefont{J.~N.} \bibnamefont{Israelachvili}},
  \emph{\bibinfo{title}{Intermolecular and Surface Forces}}
  (\bibinfo{publisher}{Academic Press}, \bibinfo{address}{London},
  \bibinfo{year}{1991}).

\bibitem[{\citenamefont{Pusey}(1991)}]{pusey:1991}
\bibinfo{author}{\bibfnamefont{P.~N.} \bibnamefont{Pusey}}, in
  \emph{\bibinfo{booktitle}{Liquids, Freezing and Glass Transition}}, edited by
  \bibinfo{editor}{\bibfnamefont{J.}~\bibnamefont{Hansen}},
  \bibinfo{editor}{\bibfnamefont{D.}~\bibnamefont{Levesque}}, \bibnamefont{and}
  \bibinfo{editor}{\bibfnamefont{J.}~\bibnamefont{Zinn-Justin}}
  (\bibinfo{publisher}{Elsevier Science Publishers B.V.}, \bibinfo{address}{Les
  Houches}, \bibinfo{year}{1991}).

\bibitem[{\citenamefont{Barker and Henderson}(1976)}]{barker:1976}
\bibinfo{author}{\bibfnamefont{J.~A.} \bibnamefont{Barker}} \bibnamefont{and}
  \bibinfo{author}{\bibfnamefont{D.}~\bibnamefont{Henderson}},
  \bibinfo{journal}{Rev. Mod. Phys.} \textbf{\bibinfo{volume}{1976}},
  \bibinfo{pages}{587} (\bibinfo{year}{1976}).

\bibitem[{\citenamefont{Knubovets et~al.}(1999)\citenamefont{Knubovets,
  Osterhout, Connolly, and Klibanov}}]{knubovets:1999}
\bibinfo{author}{\bibfnamefont{T.}~\bibnamefont{Knubovets}},
  \bibinfo{author}{\bibfnamefont{J.~J.} \bibnamefont{Osterhout}},
  \bibinfo{author}{\bibfnamefont{P.~J.} \bibnamefont{Connolly}},
  \bibnamefont{and} \bibinfo{author}{\bibfnamefont{A.~M.}
  \bibnamefont{Klibanov}}, \bibinfo{journal}{Proc. Natl. Acad. Sci. USA}
  \textbf{\bibinfo{volume}{96}}, \bibinfo{pages}{1262} (\bibinfo{year}{1999}).

\bibitem[{\citenamefont{Vagenende et~al.}(2009)\citenamefont{Vagenende, Yap,
  and Trout}}]{vagenende:2009}
\bibinfo{author}{\bibfnamefont{V.}~\bibnamefont{Vagenende}},
  \bibinfo{author}{\bibfnamefont{M.~G.~S.} \bibnamefont{Yap}},
  \bibnamefont{and} \bibinfo{author}{\bibfnamefont{B.~L.} \bibnamefont{Trout}},
  \bibinfo{journal}{Biochemistry} \textbf{\bibinfo{volume}{48}},
  \bibinfo{pages}{11084} (\bibinfo{year}{2009}).

\bibitem[{\citenamefont{Voets et~al.}(2010)\citenamefont{Voets, Cruz, Moitzi,
  Lindner, Areas, and Schurtenberger}}]{voets:2010}
\bibinfo{author}{\bibfnamefont{I.~K.} \bibnamefont{Voets}},
  \bibinfo{author}{\bibfnamefont{W.~A.} \bibnamefont{Cruz}},
  \bibinfo{author}{\bibfnamefont{C.}~\bibnamefont{Moitzi}},
  \bibinfo{author}{\bibfnamefont{P.}~\bibnamefont{Lindner}},
  \bibinfo{author}{\bibfnamefont{E.~P.~G.} \bibnamefont{Areas}},
  \bibnamefont{and}
  \bibinfo{author}{\bibfnamefont{P.}~\bibnamefont{Schurtenberger}},
  \bibinfo{journal}{J. Phys. Chem. B} \textbf{\bibinfo{volume}{114}},
  \bibinfo{pages}{11875} (\bibinfo{year}{2010}).

\bibitem[{\citenamefont{Poon et~al.}(2000)\citenamefont{Poon, Egelhaaf, Beales,
  Salonen, and Sawyer}}]{poon:2000}
\bibinfo{author}{\bibfnamefont{W.~C.~K.} \bibnamefont{Poon}},
  \bibinfo{author}{\bibfnamefont{S.~U.} \bibnamefont{Egelhaaf}},
  \bibinfo{author}{\bibfnamefont{P.~A.} \bibnamefont{Beales}},
  \bibinfo{author}{\bibfnamefont{A.}~\bibnamefont{Salonen}}, \bibnamefont{and}
  \bibinfo{author}{\bibfnamefont{L.}~\bibnamefont{Sawyer}},
  \bibinfo{journal}{J. Phys.: Condens. Matter} \textbf{\bibinfo{volume}{12}},
  \bibinfo{pages}{L569} (\bibinfo{year}{2000}).

\bibitem[{\citenamefont{Warren}(2002)}]{warren:2002}
\bibinfo{author}{\bibfnamefont{P.}~\bibnamefont{Warren}}, \bibinfo{journal}{J.
  Phys.: Condens. Matter} \textbf{\bibinfo{volume}{14}}, \bibinfo{pages}{7617}
  (\bibinfo{year}{2002}).

\bibitem[{\citenamefont{Pellicane et~al.}(2003)\citenamefont{Pellicane, Costa,
  and Caccamo}}]{pellicane:2003}
\bibinfo{author}{\bibfnamefont{G.}~\bibnamefont{Pellicane}},
  \bibinfo{author}{\bibfnamefont{D.}~\bibnamefont{Costa}}, \bibnamefont{and}
  \bibinfo{author}{\bibfnamefont{C.}~\bibnamefont{Caccamo}},
  \bibinfo{journal}{J. Phys.: Condens. Matter} \textbf{\bibinfo{volume}{15}},
  \bibinfo{pages}{375} (\bibinfo{year}{2003}).

\bibitem[{\citenamefont{Broide et~al.}(1996)\citenamefont{Broide, Tominc, and
  Saxowsky}}]{broide:1996}
\bibinfo{author}{\bibfnamefont{M.~L.} \bibnamefont{Broide}},
  \bibinfo{author}{\bibfnamefont{T.~M.} \bibnamefont{Tominc}},
  \bibnamefont{and} \bibinfo{author}{\bibfnamefont{M.~D.}
  \bibnamefont{Saxowsky}}, \bibinfo{journal}{Phys. Rev. E}
  \textbf{\bibinfo{volume}{53}}, \bibinfo{pages}{6325} (\bibinfo{year}{1996}).

\bibitem[{\citenamefont{Farnum and Zukoski}(1999)}]{farnum:1999}
\bibinfo{author}{\bibfnamefont{M.}~\bibnamefont{Farnum}} \bibnamefont{and}
  \bibinfo{author}{\bibfnamefont{C.}~\bibnamefont{Zukoski}},
  \bibinfo{journal}{Biophys. J.} \textbf{\bibinfo{volume}{76}},
  \bibinfo{pages}{2716} (\bibinfo{year}{1999}).

\bibitem[{\citenamefont{Priev et~al.}(1996)\citenamefont{Priev, Almagor,
  Yedgar, and Gavish}}]{priev:1996}
\bibinfo{author}{\bibfnamefont{A.}~\bibnamefont{Priev}},
  \bibinfo{author}{\bibfnamefont{A.}~\bibnamefont{Almagor}},
  \bibinfo{author}{\bibfnamefont{S.}~\bibnamefont{Yedgar}}, \bibnamefont{and}
  \bibinfo{author}{\bibfnamefont{B.}~\bibnamefont{Gavish}},
  \bibinfo{journal}{Biochemistry} \textbf{\bibinfo{volume}{35}},
  \bibinfo{pages}{2061} (\bibinfo{year}{1996}).

\bibitem[{\citenamefont{Sinibaldi et~al.}(2007)\citenamefont{Sinibaldi, Ortore,
  Spinozzi, Carsughi, Frielinghaus, Cinelli, Onori, and
  Mariani}}]{sinibaldi:2007}
\bibinfo{author}{\bibfnamefont{R.}~\bibnamefont{Sinibaldi}},
  \bibinfo{author}{\bibfnamefont{M.~G.} \bibnamefont{Ortore}},
  \bibinfo{author}{\bibfnamefont{F.}~\bibnamefont{Spinozzi}},
  \bibinfo{author}{\bibfnamefont{F.}~\bibnamefont{Carsughi}},
  \bibinfo{author}{\bibfnamefont{H.}~\bibnamefont{Frielinghaus}},
  \bibinfo{author}{\bibfnamefont{S.}~\bibnamefont{Cinelli}},
  \bibinfo{author}{\bibfnamefont{G.}~\bibnamefont{Onori}}, \bibnamefont{and}
  \bibinfo{author}{\bibfnamefont{P.}~\bibnamefont{Mariani}},
  \bibinfo{journal}{J. Chem. Phys.} \textbf{\bibinfo{volume}{126}},
  \bibinfo{pages}{235101} (\bibinfo{year}{2007}).

\bibitem[{\citenamefont{Merzel and Smith}(2002)}]{merzel:2002}
\bibinfo{author}{\bibfnamefont{F.}~\bibnamefont{Merzel}} \bibnamefont{and}
  \bibinfo{author}{\bibfnamefont{J.~C.} \bibnamefont{Smith}},
  \bibinfo{journal}{Proc. Natl. Acad. Sci. USA} \textbf{\bibinfo{volume}{99}},
  \bibinfo{pages}{5378} (\bibinfo{year}{2002}).

\bibitem[{\citenamefont{Esposito et~al.}(2009)\citenamefont{Esposito, Comez,
  Cinelli, Scarponi, and Onori}}]{esposito:2009}
\bibinfo{author}{\bibfnamefont{A.}~\bibnamefont{Esposito}},
  \bibinfo{author}{\bibfnamefont{L.}~\bibnamefont{Comez}},
  \bibinfo{author}{\bibfnamefont{S.}~\bibnamefont{Cinelli}},
  \bibinfo{author}{\bibfnamefont{F.}~\bibnamefont{Scarponi}}, \bibnamefont{and}
  \bibinfo{author}{\bibfnamefont{G.}~\bibnamefont{Onori}}, \bibinfo{journal}{J.
  Phys. Chem. B} \textbf{\bibinfo{volume}{113}}, \bibinfo{pages}{16420}
  (\bibinfo{year}{2009}).

\bibitem[{\citenamefont{Sedgwick et~al.}(2007)\citenamefont{Sedgwick, Cameron,
  Poon, and Egelhaaf}}]{sedgwick:2007}
\bibinfo{author}{\bibfnamefont{H.}~\bibnamefont{Sedgwick}},
  \bibinfo{author}{\bibfnamefont{J.~E.} \bibnamefont{Cameron}},
  \bibinfo{author}{\bibfnamefont{W.~C.~K.} \bibnamefont{Poon}},
  \bibnamefont{and} \bibinfo{author}{\bibfnamefont{S.~U.}
  \bibnamefont{Egelhaaf}}, \bibinfo{journal}{J. Chem. Phys.}
  \textbf{\bibinfo{volume}{127}}, \bibinfo{pages}{125102}
  (\bibinfo{year}{2007}).

\bibitem[{\citenamefont{Lu et~al.}(2003)\citenamefont{Lu, Wang, and
  Ching}}]{lu:2003}
\bibinfo{author}{\bibfnamefont{J.}~\bibnamefont{Lu}},
  \bibinfo{author}{\bibfnamefont{X.-J.} \bibnamefont{Wang}}, \bibnamefont{and}
  \bibinfo{author}{\bibfnamefont{C.-B.} \bibnamefont{Ching}},
  \bibinfo{journal}{Cryst. Growth Des.} \textbf{\bibinfo{volume}{3}},
  \bibinfo{pages}{83} (\bibinfo{year}{2003}).

\bibitem[{\citenamefont{Gosavi et~al.}(2011)\citenamefont{Gosavi, Bhamidi,
  Varanasi, and Schall}}]{gosavi:2009}
\bibinfo{author}{\bibfnamefont{R.~A.} \bibnamefont{Gosavi}},
  \bibinfo{author}{\bibfnamefont{V.}~\bibnamefont{Bhamidi}},
  \bibinfo{author}{\bibfnamefont{S.}~\bibnamefont{Varanasi}}, \bibnamefont{and}
  \bibinfo{author}{\bibfnamefont{C.~A.} \bibnamefont{Schall}},
  \bibinfo{journal}{Langmuir} \textbf{\bibinfo{volume}{25}},
  \bibinfo{pages}{4579} (\bibinfo{year}{2011}).

\bibitem[{\citenamefont{Kulkarni and Zukoski}(2002)}]{kulkarni:2002}
\bibinfo{author}{\bibfnamefont{A.~M.} \bibnamefont{Kulkarni}} \bibnamefont{and}
  \bibinfo{author}{\bibfnamefont{C.~F.} \bibnamefont{Zukoski}},
  \bibinfo{journal}{Langmuir} \textbf{\bibinfo{volume}{18}},
  \bibinfo{pages}{3090} (\bibinfo{year}{2002}).

\bibitem[{\citenamefont{Muschol and Rosenberger}(1995)}]{muschol:1995}
\bibinfo{author}{\bibfnamefont{M.}~\bibnamefont{Muschol}} \bibnamefont{and}
  \bibinfo{author}{\bibfnamefont{F.}~\bibnamefont{Rosenberger}},
  \bibinfo{journal}{J. Chem. Phys.} \textbf{\bibinfo{volume}{103}},
  \bibinfo{pages}{10424} (\bibinfo{year}{1995}).

\bibitem[{\citenamefont{Kuehner et~al.}(1997)\citenamefont{Kuehner, Heyer,
  Ramsch, Fornfeld, Blanch, and Prausnitz}}]{kuehner:1997}
\bibinfo{author}{\bibfnamefont{D.~E.} \bibnamefont{Kuehner}},
  \bibinfo{author}{\bibfnamefont{C.}~\bibnamefont{Heyer}},
  \bibinfo{author}{\bibfnamefont{C.}~\bibnamefont{Ramsch}},
  \bibinfo{author}{\bibfnamefont{U.~M.} \bibnamefont{Fornfeld}},
  \bibinfo{author}{\bibfnamefont{H.~W.} \bibnamefont{Blanch}},
  \bibnamefont{and} \bibinfo{author}{\bibfnamefont{J.~M.}
  \bibnamefont{Prausnitz}}, \bibinfo{journal}{Biophys. J.}
  \textbf{\bibinfo{volume}{73}}, \bibinfo{pages}{3211} (\bibinfo{year}{1997}).

\bibitem[{\citenamefont{Kerker}(1969)}]{kerker:1969}
\bibinfo{author}{\bibfnamefont{M.}~\bibnamefont{Kerker}},
  \emph{\bibinfo{title}{The scattering of light and other electromagnetic
  radiation}} (\bibinfo{publisher}{Acad. Press}, \bibinfo{year}{1969}).

\bibitem[{\citenamefont{Narayanan and Liu}(2003)}]{narayanan:2003}
\bibinfo{author}{\bibfnamefont{J.}~\bibnamefont{Narayanan}} \bibnamefont{and}
  \bibinfo{author}{\bibfnamefont{X.~Y.} \bibnamefont{Liu}},
  \bibinfo{journal}{Biophys. J.} \textbf{\bibinfo{volume}{84}},
  \bibinfo{pages}{523} (\bibinfo{year}{2003}).

\bibitem[{\citenamefont{Itakura et~al.}(2006)\citenamefont{Itakura, Shimada,
  Matsuyama, Saito, and Kinugasa}}]{itakura:2006}
\bibinfo{author}{\bibfnamefont{M.}~\bibnamefont{Itakura}},
  \bibinfo{author}{\bibfnamefont{K.}~\bibnamefont{Shimada}},
  \bibinfo{author}{\bibfnamefont{S.}~\bibnamefont{Matsuyama}},
  \bibinfo{author}{\bibfnamefont{T.}~\bibnamefont{Saito}}, \bibnamefont{and}
  \bibinfo{author}{\bibfnamefont{S.}~\bibnamefont{Kinugasa}},
  \bibinfo{journal}{J. Appl. Polym. Sci.} \textbf{\bibinfo{volume}{99}},
  \bibinfo{pages}{1953} (\bibinfo{year}{2006}).

\bibitem[{\citenamefont{Voet and Voet}(1990)}]{voet:1990}
\bibinfo{author}{\bibfnamefont{D.}~\bibnamefont{Voet}} \bibnamefont{and}
  \bibinfo{author}{\bibfnamefont{J.}~\bibnamefont{Voet}},
  \emph{\bibinfo{title}{Biochemistry}} (\bibinfo{publisher}{Wiley},
  \bibinfo{address}{New York}, \bibinfo{year}{1990}).

\bibitem[{\citenamefont{Stradner et~al.}(2004)\citenamefont{Stradner, Sedgwick,
  Cardinaux, Poon, Egelhaaf, and Schurtenberger}}]{stradner:2004}
\bibinfo{author}{\bibfnamefont{A.}~\bibnamefont{Stradner}},
  \bibinfo{author}{\bibfnamefont{H.}~\bibnamefont{Sedgwick}},
  \bibinfo{author}{\bibfnamefont{F.}~\bibnamefont{Cardinaux}},
  \bibinfo{author}{\bibfnamefont{W.~C.~K.} \bibnamefont{Poon}},
  \bibinfo{author}{\bibfnamefont{S.~U.} \bibnamefont{Egelhaaf}},
  \bibnamefont{and}
  \bibinfo{author}{\bibfnamefont{P.}~\bibnamefont{Schurtenberger}},
  \bibinfo{journal}{Nature} \textbf{\bibinfo{volume}{432}},
  \bibinfo{pages}{492} (\bibinfo{year}{2004}).

\bibitem[{com()}]{comment_1}
\bibinfo{note}{Similarly, Farnum and Zukoski \cite{farnum:1999} adjusted their
  values of $dn/dc_{p}$ to ensure that when using
  {Eq.}(\ref{rayleigh_ratio_b2_relation}) the extrapolation $c_{p}\to 0$
  provides the expected molar mass $M$.}

\bibitem[{\citenamefont{Russel and Benzing}(1981)}]{russel:1981}
\bibinfo{author}{\bibfnamefont{W.~B.} \bibnamefont{Russel}} \bibnamefont{and}
  \bibinfo{author}{\bibfnamefont{D.~W.} \bibnamefont{Benzing}},
  \bibinfo{journal}{J. Colloid Interface Sci.} \textbf{\bibinfo{volume}{83}},
  \bibinfo{pages}{163} (\bibinfo{year}{1981}).

\bibitem[{\citenamefont{Denton}(2000)}]{denton:2000}
\bibinfo{author}{\bibfnamefont{A.~R.} \bibnamefont{Denton}},
  \bibinfo{journal}{Phys. Rev. E} \textbf{\bibinfo{volume}{62}},
  \bibinfo{pages}{3855} (\bibinfo{year}{2000}).

\bibitem[{\citenamefont{Tanford and Roxby}(1972)}]{tanford:1972}
\bibinfo{author}{\bibfnamefont{C.}~\bibnamefont{Tanford}} \bibnamefont{and}
  \bibinfo{author}{\bibfnamefont{R.}~\bibnamefont{Roxby}},
  \bibinfo{journal}{Biochemistry} \textbf{\bibinfo{volume}{11}},
  \bibinfo{pages}{2192} (\bibinfo{year}{1972}).

\bibitem[{\citenamefont{Yin et~al.}(2003)\citenamefont{Yin, Inatomi, Wakayama,
  and Huang}}]{Yin:2003}
\bibinfo{author}{\bibfnamefont{D.~C.} \bibnamefont{Yin}},
  \bibinfo{author}{\bibfnamefont{Y.}~\bibnamefont{Inatomi}},
  \bibinfo{author}{\bibfnamefont{N.~I.} \bibnamefont{Wakayama}},
  \bibnamefont{and} \bibinfo{author}{\bibfnamefont{W.~D.} \bibnamefont{Huang}},
  \bibinfo{journal}{Cryst. Res. Technol.} \textbf{\bibinfo{volume}{38}},
  \bibinfo{pages}{785} (\bibinfo{year}{2003}).

\bibitem[{\citenamefont{Dwyer et~al.}(2000)\citenamefont{Dwyer, Gittis, Karp,
  Lattman, Spencer, Stites, and Garcia-Moreno}}]{dwyer:2000}
\bibinfo{author}{\bibfnamefont{J.~J.} \bibnamefont{Dwyer}},
  \bibinfo{author}{\bibfnamefont{A.~G.} \bibnamefont{Gittis}},
  \bibinfo{author}{\bibfnamefont{D.~A.} \bibnamefont{Karp}},
  \bibinfo{author}{\bibfnamefont{E.~E.} \bibnamefont{Lattman}},
  \bibinfo{author}{\bibfnamefont{D.~S.} \bibnamefont{Spencer}},
  \bibinfo{author}{\bibfnamefont{W.~E.} \bibnamefont{Stites}},
  \bibnamefont{and}
  \bibinfo{author}{\bibfnamefont{B.}~\bibnamefont{Garcia-Moreno}},
  \bibinfo{journal}{Biophys. J.} \textbf{\bibinfo{volume}{79}},
  \bibinfo{pages}{1610} (\bibinfo{year}{2000}).

\bibitem[{\citenamefont{Harvey and Hoekstra}(1972)}]{harvey:1972}
\bibinfo{author}{\bibfnamefont{S.~C.} \bibnamefont{Harvey}} \bibnamefont{and}
  \bibinfo{author}{\bibfnamefont{P.}~\bibnamefont{Hoekstra}},
  \bibinfo{journal}{J. Phys. Chem.} \textbf{\bibinfo{volume}{76}},
  \bibinfo{pages}{2987} (\bibinfo{year}{1972}).

\bibitem[{\citenamefont{Rashkovich et~al.}(2008)\citenamefont{Rashkovich,
  Smirnov, and Petrova}}]{rashkovich:2008}
\bibinfo{author}{\bibfnamefont{L.}~\bibnamefont{Rashkovich}},
  \bibinfo{author}{\bibfnamefont{V.}~\bibnamefont{Smirnov}}, \bibnamefont{and}
  \bibinfo{author}{\bibfnamefont{E.}~\bibnamefont{Petrova}},
  \bibinfo{journal}{Phys. Solid State} \textbf{\bibinfo{volume}{50}},
  \bibinfo{pages}{631} (\bibinfo{year}{2008}).

\bibitem[{\citenamefont{Pitera et~al.}(2001)\citenamefont{Pitera, Falta, and
  van Gunsteren}}]{pitera:2001}
\bibinfo{author}{\bibfnamefont{J.~W.} \bibnamefont{Pitera}},
  \bibinfo{author}{\bibfnamefont{M.}~\bibnamefont{Falta}}, \bibnamefont{and}
  \bibinfo{author}{\bibfnamefont{W.~F.} \bibnamefont{van Gunsteren}},
  \bibinfo{journal}{Biophys. J.} \textbf{\bibinfo{volume}{80}},
  \bibinfo{pages}{2546} (\bibinfo{year}{2001}).

\bibitem[{\citenamefont{Lide}(2005)}]{lide:2005}
\bibinfo{author}{\bibfnamefont{C.~R.} \bibnamefont{Lide}},
  \emph{\bibinfo{title}{{CRC Handbook of Chemistry and Physics}}}
  (\bibinfo{publisher}{CRC}, \bibinfo{address}{New York},
  \bibinfo{year}{2005}), \bibinfo{edition}{86th} ed.

\bibitem[{\citenamefont{{The Dow Chemical Company}}(2010)}]{dow:2010}
\bibinfo{author}{\bibnamefont{{The Dow Chemical Company}}}
  (\bibinfo{year}{2010}),
  \urlprefix\url{http://www.dow.com/glycerine/resources/}.

\bibitem[{\citenamefont{{Integrated Publishing}}(2010)}]{tpub:2010}
\bibinfo{author}{\bibnamefont{{Integrated Publishing}}} (\bibinfo{year}{2010}),
  \urlprefix\url{http://www.tpub.com/content/ArmyCRREL/SR98_02/SR98_020009.htm%
}.

\bibitem[{\citenamefont{Lu et~al.}(2009)\citenamefont{Lu, Manias, MacDonald,
  and Lanagan}}]{lu:2009}
\bibinfo{author}{\bibfnamefont{Z.}~\bibnamefont{Lu}},
  \bibinfo{author}{\bibfnamefont{E.}~\bibnamefont{Manias}},
  \bibinfo{author}{\bibfnamefont{D.~D.} \bibnamefont{MacDonald}},
  \bibnamefont{and} \bibinfo{author}{\bibfnamefont{M.}~\bibnamefont{Lanagan}},
  \bibinfo{journal}{J. Phys. Chem. A} \textbf{\bibinfo{volume}{113}},
  \bibinfo{pages}{12207} (\bibinfo{year}{2009}).

\bibitem[{\citenamefont{Nortemann et~al.}(1997)\citenamefont{Nortemann,
  Hilland, and Kaatze}}]{nortemann:1997}
\bibinfo{author}{\bibfnamefont{K.}~\bibnamefont{Nortemann}},
  \bibinfo{author}{\bibfnamefont{J.}~\bibnamefont{Hilland}}, \bibnamefont{and}
  \bibinfo{author}{\bibfnamefont{U.}~\bibnamefont{Kaatze}},
  \bibinfo{journal}{J. Phys. Chem. A} \textbf{\bibinfo{volume}{101}},
  \bibinfo{pages}{6864} (\bibinfo{year}{1997}).

\bibitem[{\citenamefont{Buchner et~al.}(1999)\citenamefont{Buchner, Hefter, and
  May}}]{buchner:1999}
\bibinfo{author}{\bibfnamefont{R.}~\bibnamefont{Buchner}},
  \bibinfo{author}{\bibfnamefont{G.~T.} \bibnamefont{Hefter}},
  \bibnamefont{and} \bibinfo{author}{\bibfnamefont{P.~M.} \bibnamefont{May}},
  \bibinfo{journal}{J. Phys. Chem. A} \textbf{\bibinfo{volume}{103}},
  \bibinfo{pages}{1} (\bibinfo{year}{1999}).

\bibitem[{\citenamefont{Gulich et~al.}(2009)\citenamefont{Gulich, Kohler,
  Lunkenheimer, and Loidl}}]{gulich:2009}
\bibinfo{author}{\bibfnamefont{R.}~\bibnamefont{Gulich}},
  \bibinfo{author}{\bibfnamefont{M.}~\bibnamefont{Kohler}},
  \bibinfo{author}{\bibfnamefont{P.}~\bibnamefont{Lunkenheimer}},
  \bibnamefont{and} \bibinfo{author}{\bibfnamefont{A.}~\bibnamefont{Loidl}},
  \bibinfo{journal}{Radiat. Environ. Biophys.} \textbf{\bibinfo{volume}{48}},
  \bibinfo{pages}{107} (\bibinfo{year}{2009}).

\bibitem[{\citenamefont{Liu et~al.}(2004)\citenamefont{Liu, Bratko, Prausnitz,
  and Blanch}}]{liu:2004}
\bibinfo{author}{\bibfnamefont{W.}~\bibnamefont{Liu}},
  \bibinfo{author}{\bibfnamefont{D.}~\bibnamefont{Bratko}},
  \bibinfo{author}{\bibfnamefont{J.~M.} \bibnamefont{Prausnitz}},
  \bibnamefont{and} \bibinfo{author}{\bibfnamefont{H.~W.}
  \bibnamefont{Blanch}}, \bibinfo{journal}{Biophys. Chem.}
  \textbf{\bibinfo{volume}{107}}, \bibinfo{pages}{289} (\bibinfo{year}{2004}).

\bibitem[{\citenamefont{Press et~al.}(1992)\citenamefont{Press, Teukolsky,
  Vetterling, and Flannery}}]{press:1992}
\bibinfo{author}{\bibfnamefont{W.~H.} \bibnamefont{Press}},
  \bibinfo{author}{\bibfnamefont{S.~A.} \bibnamefont{Teukolsky}},
  \bibinfo{author}{\bibfnamefont{W.~T.} \bibnamefont{Vetterling}},
  \bibnamefont{and} \bibinfo{author}{\bibfnamefont{B.~P.}
  \bibnamefont{Flannery}}, \emph{\bibinfo{title}{Numerical Recipes in C}}
  (\bibinfo{publisher}{Cambridge University Press}, \bibinfo{address}{New
  York}, \bibinfo{year}{1992}), \bibinfo{edition}{2nd} ed.

\bibitem[{\citenamefont{Velev et~al.}(1998)\citenamefont{Velev, Kaler, and
  Lenhoff}}]{velev:1998}
\bibinfo{author}{\bibfnamefont{O.~D.} \bibnamefont{Velev}},
  \bibinfo{author}{\bibfnamefont{E.~W.} \bibnamefont{Kaler}}, \bibnamefont{and}
  \bibinfo{author}{\bibfnamefont{A.~M.} \bibnamefont{Lenhoff}},
  \bibinfo{journal}{Biophys. J.} \textbf{\bibinfo{volume}{75}},
  \bibinfo{pages}{2682} (\bibinfo{year}{1998}).

\bibitem[{\citenamefont{Tardieu et~al.}(1999)\citenamefont{Tardieu, Le~Verge,
  Malfois, Bonnet\'e, Finet, Ri\`es-Kautt, and Belloni}}]{tardieu:1999}
\bibinfo{author}{\bibfnamefont{A.}~\bibnamefont{Tardieu}},
  \bibinfo{author}{\bibfnamefont{A.}~\bibnamefont{Le~Verge}},
  \bibinfo{author}{\bibfnamefont{M.}~\bibnamefont{Malfois}},
  \bibinfo{author}{\bibfnamefont{F.}~\bibnamefont{Bonnet\'e}},
  \bibinfo{author}{\bibfnamefont{S.}~\bibnamefont{Finet}},
  \bibinfo{author}{\bibfnamefont{M.}~\bibnamefont{Ri\`es-Kautt}},
  \bibnamefont{and} \bibinfo{author}{\bibfnamefont{L.}~\bibnamefont{Belloni}},
  \bibinfo{journal}{J. Cryst. Growth} \textbf{\bibinfo{volume}{196}},
  \bibinfo{pages}{193} (\bibinfo{year}{1999}).

\bibitem[{\citenamefont{Egelhaaf et~al.}(2004)\citenamefont{Egelhaaf, Lobaskin,
  Bauer, Merkle, and Schurtenberger}}]{egelhaaf:2004}
\bibinfo{author}{\bibfnamefont{S.~U.} \bibnamefont{Egelhaaf}},
  \bibinfo{author}{\bibfnamefont{V.}~\bibnamefont{Lobaskin}},
  \bibinfo{author}{\bibfnamefont{H.~H.} \bibnamefont{Bauer}},
  \bibinfo{author}{\bibfnamefont{H.~P.} \bibnamefont{Merkle}},
  \bibnamefont{and}
  \bibinfo{author}{\bibfnamefont{P.}~\bibnamefont{Schurtenberger}},
  \bibinfo{journal}{Eur. Phys. J. E} \textbf{\bibinfo{volume}{13}},
  \bibinfo{pages}{153} (\bibinfo{year}{2004}).

\bibitem[{\citenamefont{Hagen and Frenkel}(1994)}]{hagen:1994}
\bibinfo{author}{\bibfnamefont{M.~H.~J.} \bibnamefont{Hagen}} \bibnamefont{and}
  \bibinfo{author}{\bibfnamefont{D.}~\bibnamefont{Frenkel}},
  \bibinfo{journal}{J. Chem,. Phys.} \textbf{\bibinfo{volume}{101}},
  \bibinfo{pages}{4093} (\bibinfo{year}{1994}).

\bibitem[{\citenamefont{Cochran and Chiew}(2004)}]{cochran:2004}
\bibinfo{author}{\bibfnamefont{T.~W.} \bibnamefont{Cochran}} \bibnamefont{and}
  \bibinfo{author}{\bibfnamefont{Y.~C.} \bibnamefont{Chiew}},
  \bibinfo{journal}{J. Chem. Phys.} \textbf{\bibinfo{volume}{121}},
  \bibinfo{pages}{1480} (\bibinfo{year}{2004}).

\bibitem[{\citenamefont{Tavares et~al.}(2004)\citenamefont{Tavares, Bratko,
  Striolo, Blanch, and Prausnitz}}]{tavares:2004}
\bibinfo{author}{\bibfnamefont{F.~W.} \bibnamefont{Tavares}},
  \bibinfo{author}{\bibfnamefont{D.}~\bibnamefont{Bratko}},
  \bibinfo{author}{\bibfnamefont{A.}~\bibnamefont{Striolo}},
  \bibinfo{author}{\bibfnamefont{H.~W.} \bibnamefont{Blanch}},
  \bibnamefont{and} \bibinfo{author}{\bibfnamefont{J.~M.}
  \bibnamefont{Prausnitz}}, \bibinfo{journal}{J. Chem. Phys.}
  \textbf{\bibinfo{volume}{120}}, \bibinfo{pages}{9859} (\bibinfo{year}{2004}).

\bibitem[{\citenamefont{Bostr\"om et~al.}(2006)\citenamefont{Bostr\"om,
  Tavares, Ninham, and Prausnitz}}]{bostrom:2006}
\bibinfo{author}{\bibfnamefont{M.}~\bibnamefont{Bostr\"om}},
  \bibinfo{author}{\bibfnamefont{F.~W.} \bibnamefont{Tavares}},
  \bibinfo{author}{\bibfnamefont{B.~W.} \bibnamefont{Ninham}},
  \bibnamefont{and} \bibinfo{author}{\bibfnamefont{J.~M.}
  \bibnamefont{Prausnitz}}, \bibinfo{journal}{J. Phys. Chem. B}
  \textbf{\bibinfo{volume}{110}}, \bibinfo{pages}{24757}
  (\bibinfo{year}{2006}).

\bibitem[{\citenamefont{Lima et~al.}(2007)\citenamefont{Lima, Biscaia~{Jr.},
  Bostr\"om, Tavares, and Prausnitz}}]{lima:2007}
\bibinfo{author}{\bibfnamefont{E.~R.~A.} \bibnamefont{Lima}},
  \bibinfo{author}{\bibfnamefont{E.~C.} \bibnamefont{Biscaia~{Jr.}}},
  \bibinfo{author}{\bibfnamefont{M.}~\bibnamefont{Bostr\"om}},
  \bibinfo{author}{\bibfnamefont{F.~W.} \bibnamefont{Tavares}},
  \bibnamefont{and} \bibinfo{author}{\bibfnamefont{J.~M.}
  \bibnamefont{Prausnitz}}, \bibinfo{journal}{J. Phys. Chem. C}
  \textbf{\bibinfo{volume}{111}}, \bibinfo{pages}{16055}
  (\bibinfo{year}{2007}).

\bibitem[{\citenamefont{G\"ogelein et~al.}(2008)\citenamefont{G\"ogelein,
  N\"agele, Tuinier, Gibaud, Stradner, and Schurtenberger}}]{gogelein:2008}
\bibinfo{author}{\bibfnamefont{C.}~\bibnamefont{G\"ogelein}},
  \bibinfo{author}{\bibfnamefont{G.}~\bibnamefont{N\"agele}},
  \bibinfo{author}{\bibfnamefont{R.}~\bibnamefont{Tuinier}},
  \bibinfo{author}{\bibfnamefont{T.}~\bibnamefont{Gibaud}},
  \bibinfo{author}{\bibfnamefont{A.}~\bibnamefont{Stradner}}, \bibnamefont{and}
  \bibinfo{author}{\bibfnamefont{P.}~\bibnamefont{Schurtenberger}},
  \bibinfo{journal}{J. Chem. Phys.} \textbf{\bibinfo{volume}{129}},
  \bibinfo{pages}{085102} (\bibinfo{year}{2008}).

\bibitem[{\citenamefont{Dorsaz et~al.}(2009)\citenamefont{Dorsaz, Thurston,
  Stradner, Schurtenberger, and Foffi}}]{dorsaz:2009}
\bibinfo{author}{\bibfnamefont{N.}~\bibnamefont{Dorsaz}},
  \bibinfo{author}{\bibfnamefont{G.~M.} \bibnamefont{Thurston}},
  \bibinfo{author}{\bibfnamefont{A.}~\bibnamefont{Stradner}},
  \bibinfo{author}{\bibfnamefont{P.}~\bibnamefont{Schurtenberger}},
  \bibnamefont{and} \bibinfo{author}{\bibfnamefont{G.}~\bibnamefont{Foffi}},
  \bibinfo{journal}{J. Phys. Chem. B} \textbf{\bibinfo{volume}{113}},
  \bibinfo{pages}{1693} (\bibinfo{year}{2009}).

\bibitem[{\citenamefont{Verlet and Weis}(1972)}]{verlet:1972}
\bibinfo{author}{\bibfnamefont{L.}~\bibnamefont{Verlet}} \bibnamefont{and}
  \bibinfo{author}{\bibfnamefont{J.~J.} \bibnamefont{Weis}},
  \bibinfo{journal}{Phys. Rev. A} \textbf{\bibinfo{volume}{5}},
  \bibinfo{pages}{939} (\bibinfo{year}{1972}).

\bibitem[{\citenamefont{Wertheim}(1963)}]{wertheim:1963}
\bibinfo{author}{\bibfnamefont{M.~S.} \bibnamefont{Wertheim}},
  \bibinfo{journal}{Phys. Rev. Lett.} \textbf{\bibinfo{volume}{10}},
  \bibinfo{pages}{321} (\bibinfo{year}{1963}).

\bibitem[{\citenamefont{Throop and Bearman}(1965)}]{throop:1965}
\bibinfo{author}{\bibfnamefont{G.~J.} \bibnamefont{Throop}} \bibnamefont{and}
  \bibinfo{author}{\bibfnamefont{R.~J.} \bibnamefont{Bearman}},
  \bibinfo{journal}{J. Chem. Phys.} \textbf{\bibinfo{volume}{42}},
  \bibinfo{pages}{2408} (\bibinfo{year}{1965}).

\bibitem[{\citenamefont{Kincaid and Weis}(1977)}]{kincaid:1977}
\bibinfo{author}{\bibfnamefont{J.~M.} \bibnamefont{Kincaid}} \bibnamefont{and}
  \bibinfo{author}{\bibfnamefont{J.~J.} \bibnamefont{Weis}},
  \bibinfo{journal}{Mol. Phys.} \textbf{\bibinfo{volume}{34}},
  \bibinfo{pages}{931} (\bibinfo{year}{1977}).

\bibitem[{\citenamefont{Carnahan and Starling}(1969)}]{carnahan:1969}
\bibinfo{author}{\bibfnamefont{N.~F.} \bibnamefont{Carnahan}} \bibnamefont{and}
  \bibinfo{author}{\bibfnamefont{K.~E.} \bibnamefont{Starling}},
  \bibinfo{journal}{J. Chem. Phys.} \textbf{\bibinfo{volume}{51}},
  \bibinfo{pages}{635} (\bibinfo{year}{1969}).

\bibitem[{\citenamefont{Wood}(1952)}]{wood:1952}
\bibinfo{author}{\bibfnamefont{W.~W.} \bibnamefont{Wood}}, \bibinfo{journal}{J.
  Chem. Phys.} \textbf{\bibinfo{volume}{20}}, \bibinfo{pages}{1334}
  (\bibinfo{year}{1952}).

\bibitem[{\citenamefont{Shukla}(2000)}]{shukla:2000}
\bibinfo{author}{\bibfnamefont{K.~P.} \bibnamefont{Shukla}},
  \bibinfo{journal}{J. Chem. Phys.} \textbf{\bibinfo{volume}{112}},
  \bibinfo{pages}{10358} (\bibinfo{year}{2000}).

\bibitem[{\citenamefont{Lue and Prausnitz}(1998)}]{lue:1998}
\bibinfo{author}{\bibfnamefont{L.}~\bibnamefont{Lue}} \bibnamefont{and}
  \bibinfo{author}{\bibfnamefont{J.~M.} \bibnamefont{Prausnitz}},
  \bibinfo{journal}{J. Chem. Phys.} \textbf{\bibinfo{volume}{108}},
  \bibinfo{pages}{5529} (\bibinfo{year}{1998}).

\bibitem[{\citenamefont{Fu et~al.}(2003)\citenamefont{Fu, Li, and
  Wu}}]{fu:2003}
\bibinfo{author}{\bibfnamefont{D.}~\bibnamefont{Fu}},
  \bibinfo{author}{\bibfnamefont{Y.}~\bibnamefont{Li}}, \bibnamefont{and}
  \bibinfo{author}{\bibfnamefont{J.}~\bibnamefont{Wu}}, \bibinfo{journal}{Phys.
  Rev. E} \textbf{\bibinfo{volume}{68}}, \bibinfo{pages}{011403}
  (\bibinfo{year}{2003}).

\bibitem[{\citenamefont{Asherie et~al.}(1996)\citenamefont{Asherie, Lomakin,
  and Benedek}}]{asherie:1996}
\bibinfo{author}{\bibfnamefont{N.}~\bibnamefont{Asherie}},
  \bibinfo{author}{\bibfnamefont{A.}~\bibnamefont{Lomakin}}, \bibnamefont{and}
  \bibinfo{author}{\bibfnamefont{G.~B.} \bibnamefont{Benedek}},
  \bibinfo{journal}{Phys. Rev. Lett.} \textbf{\bibinfo{volume}{77}},
  \bibinfo{pages}{4832} (\bibinfo{year}{1996}).

\bibitem[{\citenamefont{Rosenbaum et~al.}(1996)\citenamefont{Rosenbaum, Zamora,
  and Zukoski}}]{rosenbaum:1996b}
\bibinfo{author}{\bibfnamefont{D.}~\bibnamefont{Rosenbaum}},
  \bibinfo{author}{\bibfnamefont{P.~C.} \bibnamefont{Zamora}},
  \bibnamefont{and} \bibinfo{author}{\bibfnamefont{C.~F.}
  \bibnamefont{Zukoski}}, \bibinfo{journal}{Phys. Rev. Lett.}
  \textbf{\bibinfo{volume}{76}}, \bibinfo{pages}{150} (\bibinfo{year}{1996}).

\bibitem[{\citenamefont{Muschol and Rosenberger}(1997)}]{muschol:1997}
\bibinfo{author}{\bibfnamefont{M.}~\bibnamefont{Muschol}} \bibnamefont{and}
  \bibinfo{author}{\bibfnamefont{F.}~\bibnamefont{Rosenberger}},
  \bibinfo{journal}{J. Chem. Phys.} \textbf{\bibinfo{volume}{107}},
  \bibinfo{pages}{1953} (\bibinfo{year}{1997}).

\bibitem[{\citenamefont{Rosenbaum and Zukoski}(1996)}]{rosenbaum:1996}
\bibinfo{author}{\bibfnamefont{D.~F.} \bibnamefont{Rosenbaum}}
  \bibnamefont{and} \bibinfo{author}{\bibfnamefont{C.~F.}
  \bibnamefont{Zukoski}}, \bibinfo{journal}{J. Crystal Growth}
  \textbf{\bibinfo{volume}{169}}, \bibinfo{pages}{752} (\bibinfo{year}{1996}).

\bibitem[{\citenamefont{Liu et~al.}(2005)\citenamefont{Liu, Garde, and
  Kumar}}]{liu:2005}
\bibinfo{author}{\bibfnamefont{H.}~\bibnamefont{Liu}},
  \bibinfo{author}{\bibfnamefont{S.}~\bibnamefont{Garde}}, \bibnamefont{and}
  \bibinfo{author}{\bibfnamefont{S.}~\bibnamefont{Kumar}}, \bibinfo{journal}{J.
  Chem. Phys.} \textbf{\bibinfo{volume}{123}}, \bibinfo{pages}{174505}
  (\bibinfo{year}{2005}).

\bibitem[{\citenamefont{George and Wilson}(1994)}]{george:1994}
\bibinfo{author}{\bibfnamefont{A.}~\bibnamefont{George}} \bibnamefont{and}
  \bibinfo{author}{\bibfnamefont{W.}~\bibnamefont{Wilson}},
  \bibinfo{journal}{Acta Crystallogr. D, Sect. D: Biol. Crystallogr.}
  \textbf{\bibinfo{volume}{50}}, \bibinfo{pages}{361} (\bibinfo{year}{1994}).

\bibitem[{\citenamefont{Poon}(1997)}]{poon:1997}
\bibinfo{author}{\bibfnamefont{W.~C.~K.} \bibnamefont{Poon}},
  \bibinfo{journal}{Phys. Rev. E} \textbf{\bibinfo{volume}{55}},
  \bibinfo{pages}{3762} (\bibinfo{year}{1997}).

\bibitem[{\citenamefont{Archer et~al.}(2007)\citenamefont{Archer, Pini, Evans,
  and Reatto}}]{archer:2007}
\bibinfo{author}{\bibfnamefont{A.~J.} \bibnamefont{Archer}},
  \bibinfo{author}{\bibfnamefont{D.}~\bibnamefont{Pini}},
  \bibinfo{author}{\bibfnamefont{R.}~\bibnamefont{Evans}}, \bibnamefont{and}
  \bibinfo{author}{\bibfnamefont{L.}~\bibnamefont{Reatto}},
  \bibinfo{journal}{J. Chem. Phys.} \textbf{\bibinfo{volume}{126}},
  \bibinfo{pages}{014104} (\bibinfo{year}{2007}).

\bibitem[{\citenamefont{Lomakin et~al.}(1999)\citenamefont{Lomakin, Asherie,
  and Benedek}}]{lomakin:1999}
\bibinfo{author}{\bibfnamefont{A.}~\bibnamefont{Lomakin}},
  \bibinfo{author}{\bibfnamefont{N.}~\bibnamefont{Asherie}}, \bibnamefont{and}
  \bibinfo{author}{\bibfnamefont{G.~B.} \bibnamefont{Benedek}},
  \bibinfo{journal}{Proc. Natl. Acad. Sci. USA} \textbf{\bibinfo{volume}{96}},
  \bibinfo{pages}{9465} (\bibinfo{year}{1999}).

\bibitem[{\citenamefont{G\"ogelein et~al.}()\citenamefont{G\"ogelein, Romano,
  Sciortino, and Giacometti}}]{gogelein:2011}
\bibinfo{author}{\bibfnamefont{C.}~\bibnamefont{G\"ogelein}},
  \bibinfo{author}{\bibfnamefont{F.}~\bibnamefont{Romano}},
  \bibinfo{author}{\bibfnamefont{F.}~\bibnamefont{Sciortino}},
  \bibnamefont{and}
  \bibinfo{author}{\bibfnamefont{A.}~\bibnamefont{Giacometti}},
  \emph{\bibinfo{title}{Fluid-fluid and fluid-solid transitions in the
  {K}ern-{F}renkel model from {B}arker-{H}enderson thermodynamic perturbation
  theory}}, \bibinfo{note}{submitted to The Journal of Chemical Physics}.

\bibitem[{\citenamefont{Curtis et~al.}(2002)\citenamefont{Curtis, Steinbrecher,
  Heinemann, Blanch, and Prausnitz}}]{curtis:2002}
\bibinfo{author}{\bibfnamefont{R.~A.} \bibnamefont{Curtis}},
  \bibinfo{author}{\bibfnamefont{C.}~\bibnamefont{Steinbrecher}},
  \bibinfo{author}{\bibfnamefont{M.}~\bibnamefont{Heinemann}},
  \bibinfo{author}{\bibfnamefont{H.~W.} \bibnamefont{Blanch}},
  \bibnamefont{and} \bibinfo{author}{\bibfnamefont{J.~M.}
  \bibnamefont{Prausnitz}}, \bibinfo{journal}{Biophys. Chem.}
  \textbf{\bibinfo{volume}{98}}, \bibinfo{pages}{249} (\bibinfo{year}{2002}).

\end{thebibliography}
\end{document}